\documentclass[12pt]{article}

\usepackage{color,graphicx,amsmath,amssymb,bm}

\newcommand{\A}{\ensuremath{{\cal A}}}
\newcommand{\B}{\ensuremath{{\cal B}}}
\newcommand{\C}{\ensuremath{{\cal C}}}
\newcommand{\D}{\ensuremath{{\cal D}}}
\newcommand{\G}{\ensuremath{{\cal G}}}

\title{Ferromagnetic Domain Wall and Spiral Ground States
in One-Dimensional Deformed Flat-Band Hubbard Model}

\author{Makoto Homma and Chigak Itoi
\footnote{Department of Physics, Science and Technology,
Nihon University, Tokyo, Japan}}

\begin{document}

\maketitle

\begin{abstract}
 We construct a set of exact ground states with a localized
 ferromagnetic domain wall and an extended spiral structure in a
 quasi-one-dimensional deformed flat-band Hubbard model. In the case of
 quarter filling, we show the uniqueness of the ground state with a
 fixed magnetization. The ground states with these structures are
 degenerate with the all-spin-up and all-spin-down states. This property
 of the degeneracy is the same as the domain wall solutions in the XXZ
 Heisenberg-Ising model. We derive a useful recursion relation for the
 normalization of the domain wall ground state. Using this recursion
 relation, we discuss the convergence of the ground state expectation
 values of arbitrary local operators in the infinite-volume limit. In
 the ground state of the infinite-volume system, the translational
 symmetry is spontaneously broken by this structure.  We prove that the
 cluster property holds for the domain wall ground state and excited
 states. We also estimate bounds of the ground state expectation values
 of several observables, such as one- and two-point functions of spin
 and electron number density.
 \smallskip

\noindent%
{\bf keywords}:
 ferromagnetic domain wall, spiral state, flat-band Hubbard model, exact
 solution, quantum effect, cluster property
\end{abstract}

%
%
%

\tableofcontents

\section{Introduction}

Appearance of domain structure is known to be a universal phenomenon in
ferromagnetic systems with an energy gap and with a finite correlation
length. In realistic materials, the main driving force to produce the
ferromagnetic domains is the magnetic dipole-dipole interaction, which
is much smaller than the Coulomb interaction but long-ranged.  In some
examples we introduce here, however, stable domain structures appear
without such long range interactions. Alcaraz, Salinas and Wreszinski
construct a set of ground states with a domain structure in the
ferromagnetic XXZ model with a critical boundary field in arbitrary
dimensions for an arbitrary spin \cite{ASW}. They have shown that the
degeneracy of the ground states corresponding to the location of a
domain wall center is identical to that of the ground states in the
SU(2) invariant model. In their solution, the domain wall is localized
at an arbitrary location with a finite width depending on the Ising
anisotropy parameter $(q + q^{-1})/2 > 1$ with a positive number
$q$. After their construction of the domain wall ground state, lots of
interesting properties of the solution have been discovered. Here, we
focus only on one-dimensional models and remark several interesting
results. Koma and Nachtergaele have proved the existence of the energy
gap which is independent of the location of the domain wall \cite{KN1}
in the one-dimensional XXZ model with a spin $S=1/2$. Matsui has
obtained the complete set of pure ground states in this model in the
infinite-volume limit \cite{Matsui1}. Koma and Nachtergaele have also
obtained the complete set of pure ground states with domain walls in the
one-dimensional XXZ model with an arbitrary spin $S$ in the
infinite-volume limit \cite{KN2}. Koma, Nachtergaele and Starr have
shown a singular excitation localized near the domain wall with lower
energy. Bach and Macris evaluate a spin one-point function in the domain
wall state of the one-dimensional XXZ model by a rigorous perturbation
method \cite{BM}. Datta and Kennedy discuss the stability of a domain
wall in one-dimensional quantum spin models at zero temperature by
another rigorous perturbation method Kirkwood-Thomas approach around the
Ising model as an unperturbed model \cite{DK}. In the antiferromagnetic
XXZ model, the exchange interaction destabilizes the domain structure,
while in the ferromagnetic model it preserves the domain structure at
zero temperature. The role of the quantum effects should be studied more
in many other models.

In many-electron systems, stable ferromagnetism itself is less trivial
than that in spin models. The ferromagnetism appears by purely quantum
effects in many-electron systems. There are no corresponding classical
systems to the ferromagnetic many-electron systems. Some remarkable
results for ferromagnetic ground states have recently been obtained in a
class of Hubbard models. Mielke has shown that the ground states give
saturated ferromagnetism in many-electron models on general line graphs
with special properties \cite{Mielke}. Independently, Tasaki has shown
that ground states give the saturated ferromagnetism in another class of
models \cite{Tasaki1}. These models are called the flat-band Hubbard
models, since the band structure for a single electron in the carefully
constructed hopping Hamiltonian becomes flat. Mielke and Tasaki argue
the sufficient conditions for the flat-band ferromagnetism in extended
models \cite{MT}. Tasaki has proved that the saturated ferromagnetism is
stable against a perturbation which bends the electron band
\cite{Tasaki2}. Tanaka and Ueda have shown the stability of the
saturated ferromagnetism in a more complicated two-dimensional model in
Mielke's class \cite{TU}. The ferromagnetism in flat-band Hubbard models
and their perturbed models is believed to be one universal nature of
ferromagnetism in many-electron systems. In these models, the
ferromagnetism appears in nonperturbative region in which the Coulomb
repulsion between electrons is sufficiently strong. No perturbative
expansion method can explain the ferromagnetism in these models. Only
non-perturbative methods particularly devised can work for these
problems.

In this paper, we construct a set of exact ground states in a
one-dimensional flat-band Hubbard model with coexisting domain wall and
spiral structures. We deform a flat-band Hubbard model by introducing a
complex anisotropy parameter $q$. The SU(2) spin rotation symmetry in
the original flat-band model is reduced to U(1) symmetry in our deformed
model. This anisotropy $|q| \neq 1$ leads to a localized domain wall
with a finite width and a complex $q$ leads to an extended spiral
state. We see that the domain wall ground state in our electron model
has the same degeneracy as that in the XXZ model and the same
localization property in a certain parameter regime. As in the XXZ
model, the domain structure is characterized in terms of the local order
parameter $\langle S_{x}^{(3)} \rangle$, which represents the third
component of the localized spin at site $x$. This local order parameter
takes the same sign within one domain. The domain wall center is a set
of sites $x$ defined by zeros of the local order parameter
$\langle S_{x} ^{(3)} \rangle = 0$. For complex $q=|q|e^{i \theta}$ with
real $\theta$, our electron model differs from the XXZ quantum spin
model which has no spiral ground state. The ground state in our model
has an incommensurate spiral structure with a pitch angle $\theta$. In
the case of quarter filling, we show the uniqueness of the ground state
with a fixed magnetization for an arbitrary $q$.

To calculate the ground state expectation values of observables, we
introduce a normalization function of the ground state. In particular in
the one-dimensional model, we derive a recursion relation for the
normalization function, which determines this function together with its
boundary data. This recursion relation is a linear equation in the
unknown function of a site with a function coefficient which depends on
the distance from the domain wall center. This method is similar to the
method for the normalization function in the non-product domain wall
ground states in the XXZ model obtained by Bolina, Contucci and
Nachtergaele \cite{BCN1,BCN2}. We treat this recursion relation by a
kind of perturbative expansion method in the function coefficient. This
treatment of the recursion relation enables us to discuss the
convergence of the ground state expectation values of local observables
in the infinite-volume limit. Then, we prove that the cluster property
holds for the domain wall ground state and excited states. This property
implies that the obtained representation is healthy as a quantum system
with infinite degrees of freedom. As discussed in \cite{HI}, we expect
that quantum effects in the Hubbard model stabilize the domain wall as
in the XXZ model. Finally, we show exponential decay of the truncated
correlation functions, and evaluate their decay length.

This paper is organized as follows. In section \ref{sec:def_of_model},
we define a deformed flat-band model on a quasi-one-dimensional
lattice. We give an explicit form of a Hamiltonian. In section
\ref{sec:GS}, we construct ground states, and prove the uniqueness of
the ground state with a fixed magnetization. We also introduce different
basis of ground states whose state is superposed over all magnetizations
with certain weights. In section \ref{sec:N-func}, we define a
normalization function and extract its convergent factor, which gives us
good representations of the expectation values of arbitrary local
operators in the ground state. We obtain a recursion relation for the
normalization function which is useful to prove important properties of
the observables. In section \ref{sec:infinite}, we evaluate the
expectation values of local operators in the infinite-volume limit. We
prove the cluster property of the domain wall ground state on the basis
of the recursion relation.

\section{The Model \label{sec:def_of_model}}

\subsection{Definition of the Model}

We define a one-dimensional lattice $\Lambda$ with a ``length'' $L$ by
\begin{equation}
 \Lambda :=
 \left\{
  x = \frac{y}{2} ~ \Bigl| ~ y \in {\mathbb Z} \mbox{ with }
  |y| \leq L
 \right\},
\end{equation}
where $L=2l+1$ is an odd integer. $\Lambda$ can be decomposed two sub
lattices $\Lambda_{o}$ and $\Lambda^{\prime}$ defined by
\begin{equation}
 \Lambda_{o} :=
 \left\{
  x \in \Lambda ~ | ~ x \mbox{ is an integer}
 \right\}
\end{equation}
and
\begin{equation}
 \Lambda^{\prime} :=
 \left\{
  x \in \Lambda ~ | ~ x \mbox{ is a half-odd-integer}
 \right\}.
\end{equation}
\begin{figure}[tb]
 \begin{center}
 \setlength{\unitlength}{1mm}
 \begin{picture}(100,20)(-50,-10)
  \put(-40,-10){\circle{3}}
  \put(-20,-10){\circle{3}}
  \put(0,-10){\circle{3}}
  \put(20,-10){\circle{3}}
  \put(40,-10){\circle{3}}
  \put(-30,0){\circle*{2}}
  \put(-10,0){\circle*{2}}
  \put(10,0){\circle*{2}}
  \put(30,0){\circle*{2}}
  \put(-41.5,-10){\line(-1,0){4}}
  \put(-38.5,-10){\line(1,0){17}}
  \put(-18.5,-10){\line(1,0){17}}
  \put(1.5,-10){\line(1,0){17}}
  \put(21.5,-10){\line(1,0){17}}
  \put(41.5,-10){\line(1,0){4}}
  \put(-41,-9){\line(-1,1){4}}
  \put(-39,-9){\line(1,1){8.5}}
  \put(-21,-9){\line(-1,1){8.5}}
  \put(-19,-9){\line(1,1){8.5}}
  \put(-1,-9){\line(-1,1){8.5}}
  \put(1,-9){\line(1,1){8.5}}
  \put(19,-9){\line(-1,1){8.5}}
  \put(21,-9){\line(1,1){8.5}}
  \put(39,-9){\line(-1,1){8.5}}
  \put(41,-9){\line(1,1){4}}
 \end{picture}
 \end{center}
 \caption{The lattice $\Lambda$. White circles are sites in
 $\Lambda_{o}$ and black dots are sites in $\Lambda^{\prime}$. Electrons
 at a site can hop to other site if the site is connected to the
 original site with a line.}
 \label{fig:lattice}
\end{figure}
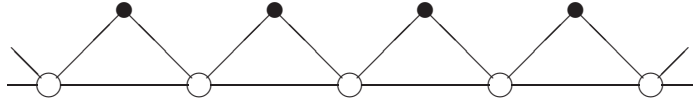

We define a deformed flat-band Hubbard model on the lattice
$\Lambda$. We denote creation and annihilation operators by
$c_{x, \sigma}^{\dagger}$ and $c_{x, \sigma}$ respectively, for an
electron at site $x \in \Lambda$ with spin
$\sigma = \uparrow, \downarrow$. They satisfy the standard
anticommutation relations
\begin{equation}
 \{ c_{x, \sigma}, c_{y, \tau}^{\dagger} \} =
 \delta_{x, y}\delta_{\sigma, \tau},~~
 \{ c_{x, \sigma}, c_{y, \tau} \} =
 \{ c_{x, \sigma}^{\dagger}, c_{y, \tau}^{\dagger} \} =
 0
 \label{eq:anticommutator_for_c}
\end{equation}
for $x, y \in \Lambda$ and $\sigma, \tau = \uparrow, \downarrow$,
where $\{A, B\} = A B + B A$. The electron number operator is
defined by
\begin{equation}
 n_{x, \sigma} := c_{x, \sigma}^{\dagger} c_{x, \sigma},
\end{equation}
whose eigenvalue gives the number of electron with spin $\sigma$ at site
$x$. In the present paper, we use the open boundary condition. This
condition is given by defining $c_{x, \sigma} = 0$ for $|x| > L/2$. We
consider a Hubbard model with a Hamiltonian
\begin{equation}
 H = H_{\rm hop} + H_{\rm int}.
  \label{eq:H_tot}
\end{equation}
The interaction Hamiltonian is given by
\begin{equation}
 H_{\rm int} = U \sum_{x \in \Lambda}
 n_{x, \uparrow} n_{x, \downarrow}
 \label{eq:H_int}
\end{equation}
with a positive number $U$ which represents an on-site Coulomb repulsion
energy. The hopping Hamiltonian is defined by
\begin{align}
 H_{\rm hop} =
 t \sum_{\sigma = \uparrow, \downarrow}
 \sum_{x \in \Lambda^{\prime}}
 \left[
  (q^{-p(\sigma)/4})^{\ast} c_{x - \frac{1}{2}, \sigma}^{\dagger}
  + \lambda c_{x, \sigma}^{\dagger}
  + (q^{p(\sigma)/4})^{\ast} c_{x + \frac{1}{2}, \sigma}^{\dagger}
 \right]
 \nonumber \\
 \times
 \left[
  q^{-p(\sigma)/4} c_{x - \frac{1}{2}, \sigma}
  + \lambda c_{x, \sigma}
  + q^{p(\sigma)/4} c_{x + \frac{1}{2}, \sigma}
 \right],
 \label{eq:H_{hop}}
\end{align}
where $p(\sigma) = 1$ if $\sigma = \uparrow$ and $p(\sigma) = -1$ if
$\sigma = \downarrow$ with $t > 0$.
We can represent the hopping Hamiltonian also in a standard form
with a hopping matrix $t_{x,y}^{(\sigma)}$
\begin{equation}
 H_{\rm hop} =
 \sum_{\sigma = \uparrow, \downarrow} \sum_{x, y \in \Lambda}
 t_{x, y}^{(\sigma)} c_{x, \sigma}^{\dagger} c_{y, \sigma}
 \label{eq:H_hop2}
\end{equation}
where
\begin{equation}
 t_{x, y}^{(\sigma)}/t = (t_{y, x}^{(\sigma)})^{\ast}/t =
 \begin{cases}
  |q|^{\frac{1}{2}} + |q|^{-\frac{1}{2}} &
  \mbox{if } x = y \in \Lambda_{o}\\
  \lambda^{2} &
  \mbox{if } x = y \in \Lambda^{\prime}\\
  \lambda q^{-p(\sigma)/4} &
  \mbox{if } x = y + \frac{1}{2} \mbox{ and } y \in \Lambda_{o}\\
  \lambda q^{p(\sigma)/4} &
  \mbox{if } x = y - \frac{1}{2} \mbox{ and } y \in \Lambda_{o}\\
  (q/ q^{\ast})^{p(\sigma)/4} &
  \mbox{if } x = y + 1 \in \Lambda_{o}\\
  0 & \mbox{otherwise}
 \end{cases}
 .
\end{equation}
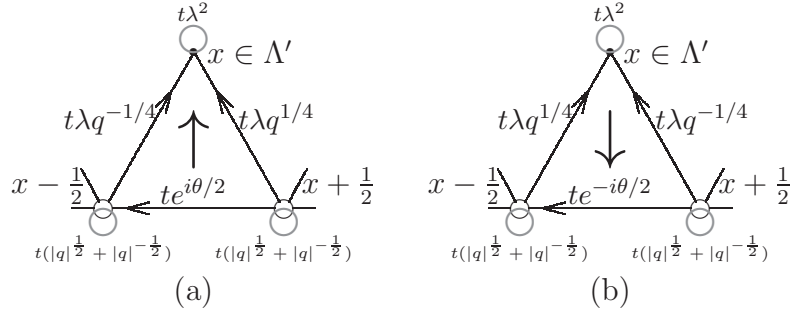
\begin{figure}[tb]
 \begin{center}
 \setlength{\unitlength}{6mm}%
 \begin{picture}(8.5,7)(-4,-2)
  \put(0,1.53){\makebox(0,0){\bf\Huge$\uparrow$}}
  \put(0,3.46){\circle*{0.2}}
  \put(-2,0){\circle{0.4}}
  \put(2,0){\circle{0.4}}
  \put(-2.2,0){\line(-1,0){0.5}}
  \put(-1.8,0){\line(1,0){3.6}}
  \put(2.2,0){\line(1,0){0.5}}
  \qbezier[200](-1.89,0.15)(-1,1.73)(0,3.46)
  \qbezier[200](1.89,0.15)(1,1.73)(0,3.46)
  \qbezier[50](-2.11,0.15)(-2.25,0.435)(-2.5,0.87)
  \qbezier[50](2.11,0.15)(2.25,0.435)(2.5,0.87)
  \qbezier[20](-0.500,2.598)(-0.565,2.357)(-0.629,2.115)
  \qbezier[20](-0.500,2.598)(-0.677,2.421)(-0.853,2.245)
  \qbezier[20](0.500,2.598)(0.565,2.357)(0.629,2.115)
  \qbezier[20](0.500,2.598)(0.677,2.421)(0.853,2.245)
  \qbezier[20](-1.5,0)(-1.259,0.065)(-1.017,0.129)
  \qbezier[20](-1.5,0)(-1.259,-0.065)(-1.017,-0.129)
  \thicklines
  \put(0,3.76){\color[named]{Gray} \circle{0.6}}
  \put(-2,-0.3){\color[named]{Gray} \circle{0.6}}
  \put(2,-0.3){\color[named]{Gray} \circle{0.6}}
  \thinlines
  \put(0.3,3.46){\makebox(0,0)[l]{$x \in \Lambda^{\prime}$}}
  \put(-2.4,0.1){\makebox(0,0)[rb]{$x-\frac{1}{2}$}}
  \put(2.4,0.1){\makebox(0,0)[lb]{$x+\frac{1}{2}$}}
  \put(-0.8,2){\makebox(0,0)[r]{\small$t\lambda q^{-1/4}$}}
  \put(1,2){\makebox(0,0)[l]{\small$t\lambda q^{1/4}$}}
  \put(0,0.1){\makebox(0,0)[b]{\small$t e^{i \theta /2}$}}
  \put(-2,-0.6){\makebox(0,0)[t]
  {\tiny $t(|q|^{\frac{1}{2}}+|q|^{-\frac{1}{2}})$}}
  \put(2,-0.6){\makebox(0,0)[t]
  {\tiny $t(|q|^{\frac{1}{2}}+|q|^{-\frac{1}{2}})$}}
  \put(0,4.16){\makebox(0,0)[b]
  {\scriptsize $t\lambda^2$}}
  \put(0,-1.5){\makebox(0,0)[t]{(a)}}
 \end{picture}
 \begin{picture}(8.5,7)(-4.5,-2)
  \put(0,1.53){\makebox(0,0){\bf\Huge$\downarrow$}}
  \put(0,3.46){\circle*{0.2}}
  \put(-2,0){\circle{0.4}}
  \put(2,0){\circle{0.4}}
  \put(-2.2,0){\line(-1,0){0.5}}
  \put(-1.8,0){\line(1,0){3.6}}
  \put(2.2,0){\line(1,0){0.5}}
  \qbezier[200](-1.89,0.15)(-1,1.73)(0,3.46)
  \qbezier[200](1.89,0.15)(1,1.73)(0,3.46)
  \qbezier[50](-2.11,0.15)(-2.25,0.435)(-2.5,0.87)
  \qbezier[50](2.11,0.15)(2.25,0.435)(2.5,0.87)
  \qbezier[20](-0.500,2.598)(-0.565,2.357)(-0.629,2.115)
  \qbezier[20](-0.500,2.598)(-0.677,2.421)(-0.853,2.245)
  \qbezier[20](0.500,2.598)(0.565,2.357)(0.629,2.115)
  \qbezier[20](0.500,2.598)(0.677,2.421)(0.853,2.245)
  \qbezier[20](-1.5,0)(-1.259,0.065)(-1.017,0.129)
  \qbezier[20](-1.5,0)(-1.259,-0.065)(-1.017,-0.129)
  \thicklines
  \put(0,3.76){\color[named]{Gray} \circle{0.6}}
  \put(-2,-0.3){\color[named]{Gray} \circle{0.6}}
  \put(2,-0.3){\color[named]{Gray} \circle{0.6}}
  \thinlines
  \put(0.3,3.46){\makebox(0,0)[l]{$x \in \Lambda^{\prime}$}}
  \put(-2.4,0.1){\makebox(0,0)[rb]{$x-\frac{1}{2}$}}
  \put(2.4,0.1){\makebox(0,0)[lb]{$x+\frac{1}{2}$}}
  \put(-0.8,2){\makebox(0,0)[r]{\small$t\lambda q^{1/4}$}}
  \put(1,2){\makebox(0,0)[l]{\small$t\lambda q^{-1/4}$}}
  \put(0,0.1){\makebox(0,0)[b]{\small$t e^{-i \theta /2}$}}
  \put(-2,-0.6){\makebox(0,0)[t]
  {\tiny $t(|q|^{\frac{1}{2}}+|q|^{-\frac{1}{2}})$}}
  \put(2,-0.6){\makebox(0,0)[t]
  {\tiny $t(|q|^{\frac{1}{2}}+|q|^{-\frac{1}{2}})$}}
  \put(0,4.16){\makebox(0,0)[b]
  {\scriptsize $t\lambda^2$}}
  \put(0,-1.5){\makebox(0,0)[t]{(b)}}
 \end{picture}
 \end{center}
 \caption{Arrows and gray circles depict hopping amplitudes and on-site
 potentials respectively for electrons with up(a) and down(b) spins on
 a unit cell. For complex $q=|q| e^{i \theta}$, the hopping amplitude
 opposite to the arrow is the complex conjugate of that along the
 arrow.}
 \label{fig:hopping-diagram}
\end{figure}

The Hilbert space of the model is spanned by the following basis
\begin{equation}
 \Phi_{\A, \B} =
 \left(
  \prod_{x \in \A} c_{x, \uparrow}^{\dagger}
 \right)
 \left(
  \prod_{y \in \B} c_{y, \downarrow}^{\dagger}
 \right)
 \Phi_{\rm vac} \label{eq:basis_of_H},
\end{equation}
where $\A, \B$ are subsets of $\Lambda$ and $\Phi_{\rm vac}$
is the unique normalized state with no electrons defined by
$c_{x, \sigma} \Phi_{\rm vac} = 0$ for any $x \in \Lambda$ and
$\sigma = \uparrow, \downarrow$.
In the present work, we consider the Hilbert space ${\cal H}$ with the
electron number fixed to $L = |\Lambda_{o}|$, which is spanned by the
basis states (\ref{eq:basis_of_H}) with $|\A| + |\B| = L$.%
\footnote{Throughout this paper, $|\A|$ denotes the number of all
elements in a set $\A$.}

\subsection{Symmetries of the Model Hamiltonian}

Let us discuss the symmetries in our model.
The local spin operator $S_{x}^{(j)}$ and the total-spin operator
$S_{\rm tot}^{(j)}$ are defined by
\begin{equation}
 S_{x}^{(j)} =
 \frac{1}{2} \sum_{\sigma, \tau}
 c_{x, \sigma}^{\dagger} ({\cal P}^{(j)})_{\sigma, \tau} c_{x, \tau}
\end{equation}
and
\begin{equation}
 S_{\rm tot}^{(j)} = \sum_{x \in \Lambda} S_{x}^{(j)}
\end{equation}
with $j = 1, 2, 3$, where ${\cal P}^{(j)}$ is the Pauli matrix
\begin{equation}
 {\cal P}^{(1)} =
 \begin{pmatrix}
  0 & 1\\ 1 & 0
 \end{pmatrix}
 ,~
 {\cal P}^{(2)} =
 \begin{pmatrix}
  0 & -i\\ i & 0
 \end{pmatrix}
 ,~
 {\cal P}^{(3)} =
 \begin{pmatrix}
 1 & 0 \\ 0 & -1
 \end{pmatrix}
 .
\end{equation}

The first important symmetry is U(1) symmetry. The Hamiltonian commutes
with $S_{\rm tot}^{(3)}$, and an eigenvalue of $S_{\rm tot}^{(3)}$ is
called a magnetization which takes $\pm 1/2$, $\pm 3/2$, $\cdots$,
$\pm L/2$. We define U(1) transformation by a following map
$U_{\varphi}$ which is generated by $S_{\rm tot}^{(3)}$ in a finite
system:
\begin{equation}
 U_{\varphi}(c_{x, \sigma}^{\dagger}) =
  e^{i \frac{\varphi}{2} p(\sigma)} c_{x, \sigma}^{\dagger}
  \quad \mbox{and} \quad
  U_{\varphi}(c_{x, \sigma}) =
  e^{- i \frac{\varphi}{2} p(\sigma)} c_{x, \sigma},
\label{U}
\end{equation}
where $\varphi$ is a real parameter. This map is linear
\begin{equation}
 U_{\varphi}(A + B) = U_{\varphi}(A) + U_{\varphi}(B)
\end{equation}
and homomorphic
\begin{equation}
U_{\varphi}(A B) = U_{\varphi}(A) U_{\varphi}(B),
\end{equation}
for arbitrary operators $A$ and $B$. Each spin operator is transformed
to the following form under the U(1) transformation:
\begin{align}
 U_{\varphi}(S_{x}^{(1)}) = &
 \cos \varphi ~ S_{x}^{(1)} - \sin \varphi ~ S_{x}^{(2)},
 \\
 U_{\varphi}(S_{x}^{(2)}) = &
 \sin \varphi ~ S_{x}^{(1)} + \cos \varphi ~ S_{x}^{(2)},
 \\
 U_{\varphi}(S_{x}^{(3)}) = & S_{x}^{3}.
\end{align}
Note that the U(1) symmetry is enhanced to the SU(2) in the case of
$q = 1$, i.e. the Hamiltonian commutes with the all total-spin operators
$S_{\rm tot}^{(j)}$ $(j = 1, 2, 3)$.

The second important symmetry is a ${\mathbb Z}_2$ symmetry defined by a
linear and homomorphic map $\Pi$:
\begin{equation}
 \Pi( c_{x, \sigma} ) = c_{-x, \overline{\sigma}}
  \quad \mbox{and} \quad
 \Pi( c_{x, \sigma}^{\dagger} ) = c_{-x, \overline{\sigma}}^{\dagger}
\label{eq:pi}
\end{equation}
where $\overline{\sigma}$ takes $\downarrow$ if $\sigma = \uparrow$ and
$\uparrow$ if $\sigma = \downarrow$.
Note the ${\mathbb Z}_2$ transformation of the total spin operator
$\Pi(S_{\rm tot}^{(3)}) = - S_{\rm tot}^{(3)}$. An energy eigenstate
with a magnetization $M$ is transformed by $\Pi$ into another eigenstate
with a total magnetization $-M$, which belongs to the same energy
eigenvalue.

The Hamiltonian of a system with a large volume
is invariant under a translation
\begin{equation}
 T_{u} (c_{x, \sigma}) = c_{x + u, \sigma}
  \quad \mbox{and} \quad
 T_{u} (c_{x, \sigma}^{\dagger}) = c_{x + u, \sigma}^{\dagger},
\label{T}
\end{equation}
except several terms near the boundaries, where $u$ is an integer.
Thus, the translational symmetry appears after taking the
infinite-volume limit.

\section{Ground State of the Model \label{sec:GS}}

In this section, we construct ground states of the deformed
flat-band Hubbard model.
The construction method is based on the original method obtained
by Mielke and Tasaki
for the positive semi-definite Hamiltonian in
the flat-band Hubbard model \cite{Mielke,Tasaki1}.

\subsection{Localized Electron Operators}

We define localized electron operators $a_{x, \sigma}^{\dagger}$ and
$d_{x, \sigma}$ by
\begin{equation}
 a_{x, \sigma}^{\dagger} =
 \begin{cases}
  \displaystyle
  - q^{p(\sigma)/4} c_{x - \frac{1}{2}, \sigma}^{\dagger}
  + \lambda c_{x, \sigma}^{\dagger}
  - q^{-p(\sigma)/4} c_{x + \frac{1}{2}, \sigma}^{\dagger}
  & \mbox{if } x \in \Lambda_{o}\\
  \displaystyle
  \frac{1}{\lambda} c_{x, \sigma}^{\dagger}
  & \mbox{if } x \in \Lambda^{\prime}
 \end{cases}
 , \label{eq:def_of_a}
\end{equation}
and
\begin{equation}
 d_{x, \sigma} =
 \begin{cases}
  \displaystyle
  \frac{1}{\lambda} c_{x, \sigma}
  & \mbox{if } x \in \Lambda_{o}\\
  \displaystyle
  q^{-p(\sigma)/4} c_{x - \frac{1}{2}, \sigma}
  + \lambda c_{x, \sigma}
  + q^{p(\sigma)/4} c_{x + \frac{1}{2}, \sigma}
  & \mbox{if } x \in \Lambda^{\prime}
 \end{cases}
 . \label{eq:def_of_d}
\end{equation}
$c_{x, \sigma}^{\dagger}$ and $c_{x, \sigma}$ can be written in terms of
$a_{x, \sigma}^{\dagger}$ and $d_{x, \sigma}$:
\begin{equation}
 c_{x, \sigma}^{\dagger} =
 \begin{cases}
  \displaystyle
  q^{p(\sigma)/4} a_{x - \frac{1}{2}, \sigma}^{\dagger}
  + \frac{1}{\lambda} a_{x, \sigma}^{\dagger}
  + q^{-p(\sigma)/4} a_{x + \frac{1}{2}, \sigma}^{\dagger}
  & \mbox{if } x \in \Lambda_{0}\\
  \lambda a_{x, \sigma}^{\dagger}
  & \mbox{if } x \in \Lambda^{\prime}
 \end{cases}
 , \label{eq:inv_of_cd}
\end{equation}
and
\begin{equation}
 c_{x, \sigma} =
 \begin{cases}
  \lambda d_{x, \sigma}
  & \mbox{if } x \in \Lambda_{o}\\
  \displaystyle
  - q^{-p(\sigma)/4} d_{x - \frac{1}{2}, \sigma}
  + \frac{1}{\lambda} d_{x, \sigma}
  - q^{p(\sigma)/4} d_{x + \frac{1}{2}, \sigma}
  & \mbox{if } x \in \Lambda^{\prime}
 \end{cases}
 . \label{eq:inv_of_c}
\end{equation}
For an open boundary condition, we define $c_{x, \sigma} = 0$ for
$|x| > L/2$ which means $a_{x, \sigma}^{\dagger} = 0$ and
$d_{x, \sigma} = 0$ for $|x| > L/2$. The
relations (\ref{eq:def_of_a}), (\ref{eq:def_of_d}), (\ref{eq:inv_of_cd})
and (\ref{eq:inv_of_c}) are also valid consistently under this condition.
The inverse relations (\ref{eq:inv_of_cd})
and (\ref{eq:inv_of_c}) implies that the Hilbert space ${\cal H}$
can be spanned also by the basis states
\begin{equation}
 \Phi_{\A, \B}^{\prime} =
 \left(
  \prod_{x \in \A} a_{x, \uparrow}^{\dagger}
 \right)
 \left(
  \prod_{y \in \B} a_{y, \downarrow}^{\dagger}
 \right)
 \Phi_{\rm vac}
\end{equation}
where $\A$ and $\B$ are subsets of $\Lambda$ with
$|\A| + |\B| = L$.
Note that $a_{x, \sigma}^{\dagger}$ and $d_{x, \sigma}$ satisfy the
canonical anticommutation relations
\begin{equation}
 \{ a_{x, \sigma}^{\dagger}, d_{y, \tau}\} = \delta_{x, y}
 \delta_{\sigma, \tau},~~
 \{ a_{x, \sigma}^{\dagger}, a_{y, \tau}^{\dagger}\} =
 \{ d_{x, \sigma}, d_{y, \tau} \} =0.
 \label{eq:anticommutator_for_a_and_d}
\end{equation}
The hopping Hamiltonian can be written in terms of $d_{x, \sigma}$,
\begin{equation}
 H_{\rm hop} =
 \sum_{\sigma = \uparrow, \downarrow} \sum_{x \in \Lambda^{\prime}}
 d_{x, \sigma}^{\dagger}d_{x, \sigma}.
\end{equation}
This form of the hopping Hamiltonian indicates the positive
semi-definiteness $H_{\rm hop} \geq 0$. The positive semi-definiteness
of the interaction Hamiltonian $H_{\rm int} \geq 0$ is also clear
because $n_{x,\sigma} = c_{x, \sigma}^{\dagger} c_{x, \sigma} \geq 0$.
Thus, the total Hamiltonian is also positive semi-definite
\begin{equation}
H=H_{\rm hop}+H_{\rm int} \geq 0.
\label{psd}
\end{equation}
First, we define all-spin-up and all-spin-down states by
\begin{equation}
 \Phi_{\uparrow} =
 \left(
  \prod_{x \in \Lambda_{o}} a_{x, \uparrow}^{\dagger}
 \right) \Phi_{\rm vac}
 \quad \mbox{and} \quad
 \Phi_{\downarrow} =
 \left(
  \prod_{x \in \Lambda_{o}} a_{x, \downarrow}^{\dagger}
 \right)
 \Phi_{\rm vac}.
\label{allup}
\end{equation}
We easily verify $H \Phi_{\uparrow} = 0=H \Phi_{\downarrow}$ from the
anticommutation relations of $a_{x, \sigma}^{\dagger}$ and
$d_{x, \sigma}$, and therefore these states are ground states of $H$.

Next, we determine all other ground states.
The conditions on a state $\Phi$ to be a ground state
$H_{\rm hop} \Phi = 0$ and $H_{\rm int} \Phi = 0$ are equivalent to
\begin{equation}
 d_{x, \sigma} \Phi = 0 \label{eq:cond-of-GS-1}
\end{equation}
for all $x \in \Lambda^{\prime}$ and
\begin{equation}
 c_{y, \uparrow} c_{y, \downarrow} \Phi = 0 \label{eq:cond-of-GS-2}
\end{equation}
for all $y \in \Lambda$ with $\sigma = \uparrow, \downarrow$.

We consider a restricted Hilbert space ${\cal H}_{M}$ spanned by the
states with a magnetization $M$. Since the magnetization is a good quantum
number, we find other ground states in each restricted sub-space
${\cal H}_{M}$.
Let $\Phi(M)$ be a ground state with a magnetization $M$. We expand
$\Phi(M)$ into the following series
\begin{equation}
 \Phi(M) = \sum_{\A, \B}
 \psi(\A, \B)
 \left(
  \prod_{x \in \A} a_{x, \uparrow}^{\dagger}
 \right)
 \left(
  \prod_{y \in \B} a_{y, \downarrow}^{\dagger}
 \right)
 \Phi_{\rm vac} \label{eq:exp-of-GS-1},
\end{equation}
where the sum is taken over all subsets
$\A, \B \subset \Lambda$ with
$|\A| = (L + 2 M)/2 $ and $|\B| = (L - 2 M)/2$, and
$\psi(\A, \B)$ is a
coefficient. The first condition (\ref{eq:cond-of-GS-1}) implies that
$\psi(\A, \B)$ does not vanish only for
$\A\ \cap \Lambda^{\prime} = \emptyset$ and
$\B \cap \Lambda^{\prime} = \emptyset$. The second condition
(\ref{eq:cond-of-GS-2}) for $y \in \Lambda_{o}$ becomes
\begin{equation}
 d_{y, \uparrow} d_{y, \downarrow} \Phi(M) = 0,
\end{equation}
due to the definition of $d_{y, \sigma}$. To satisfy this condition,
$\psi(\A, \B)$ must vanish for
$\A \cap \B \neq \emptyset$. Thus the expansion
(\ref{eq:exp-of-GS-1}) can be written in the form:
\begin{equation}
 \Phi(M) =
 \sum_{\sigma \in {\cal S}_{M}} \phi(\sigma)
 \prod_{x \in \Lambda_{o}} a_{x, \sigma_{x}}^{\dagger} \Phi_{\rm vac}
\end{equation}
where ${\cal S}_{M}$ is a set of all possible spin configurations
$\sigma = (\sigma_{x})_{x \in \Lambda_{o}}$ with a magnetization $M$:
\begin{equation}
 {\cal S}_{M} =
 \left\{
  \sigma = (\sigma_{x})_{x \in \Lambda_{o}}
  ~ \Biggl| ~
  \frac{1}{2}\sum_{x \in \Lambda_{o}} (-1)^{\frac{1 - p(\sigma_{x})}{2}}
  = M, ~~
  \sigma_{x} = \uparrow, \downarrow
 \right\}.
\end{equation}
In this representation, the second condition (\ref{eq:cond-of-GS-2}) for
$y \in \Lambda^{\prime}$ is equivalent to
\begin{equation}
 \left(
 d_{x, \uparrow} d_{x + 1, \downarrow}
 - q d_{x, \downarrow} d_{x + 1, \uparrow}
 \right)
 \Phi = 0
\end{equation}
where we use inverse relations (\ref{eq:inv_of_c}) and anticommutation
relations (\ref{eq:anticommutator_for_a_and_d}) with
$x = y - \frac{1}{2} \in \Lambda_{o}$.
We find that this condition is satisfied if and only if the coefficients
satisfy
\begin{equation}
 \phi(\sigma) = q^{[p(\sigma_{x}) - p(\sigma_{x + 1})]/2}
 \phi(\sigma_{x, x + 1})
 \label{eq:cond_for_phi}
\end{equation}
for $x \in \Lambda_{o}$ with $x < (L - 3)/2$, where $\sigma_{x, x + 1}$
is a spin configuration obtained by the exchange $\sigma_{x}$ and
$\sigma_{x + 1}$ in the original configuration $\sigma$.
This relation implies the uniqueness of the ground state with a fixed
magnetization, since two arbitrary spin configurations can be
related by successive exchanges of two nearest neighbour spins.
Therefore the degeneracy of those ground states is exactly the same as
that in the SU(2) symmetric model, as in the domain wall ground states
in the XXZ model \cite{ASW}.

If we define $S_{q}^{-}$ by
\begin{equation}
 S_{q}^{-} := \sum_{x \in \Lambda} q^{x}
 a_{x, \downarrow}^{\dagger} d_{x, \uparrow}.
\end{equation}
The ground state with a magnetization $M$ is obtained
by acting this operator $L - 2 M$ times on the all-spin-up state
\begin{equation}
 \Phi(M) = \left( S_{q}^{-} \right)^{L - 2 M} \Phi_{\uparrow}.
\end{equation}

\subsection{Another Set of Ground States}

To explore the property of the ground state, we write it in a
more explicit form as obtained by Gottstein and Werner in \cite{GW}.
We define the ground state $\Psi(\zeta)$ by
\begin{equation}
 \Psi(\zeta) :=
 \left(
  \prod_{x \in \Lambda_{0}}
  \alpha_{x}^{\dagger}(\zeta)
 \right)
 \Phi_{\rm vac}
 = \sum_{n = 0}^{L} \zeta^{n} (S_{q}^{-})^{n} \Phi_{\uparrow}
\end{equation}
where
$\alpha_{x}^{\dagger}(\zeta) := a_{x, \uparrow}^{\dagger}
+ \zeta q^{x} a_{x, \downarrow}^{\dagger}$. We fix the order of the
electron operators depending on their sites.
In the XXZ model, the state which corresponds to this ground state $\Psi(\zeta)$ is a
pure state in the infinite-volume limit, and therefore we can expect that this
state is stable in the infinite-volume limit \cite{Matsui2}.
Note that the state $\Psi(\zeta)$ defined here in the Hubbard
model is not a product state unlike the corresponding state in the XXZ
model.

\subsection{Spin One-Point Functions}

A spin expectation value of a localized spin at site $x$ in the
ground state $\Psi(\zeta)$ is written as
\begin{equation}
 \langle S_{x}^{(j)} \rangle_{\zeta}
 =
 \sum_{\sigma, \tau = \uparrow, \downarrow}
 \frac{({\cal P}^{(j)})_{\sigma, \tau}}{2}
 \frac{(\Psi(\zeta), c_{x, \sigma}^{\dagger} c_{x, \tau} \Psi(\zeta))}
 {\| \Psi(\zeta) \|^{2}}
 =
 \sum_{\sigma, \tau = \uparrow, \downarrow}
 \frac{({\cal P}^{(j)})_{\sigma, \tau}}{2}
 \frac{(c_{x, \sigma} \Psi(\zeta), c_{x, \tau} \Psi(\zeta))}
 {\| \Psi(\zeta) \|^{2}}.
\end{equation}
The following anticommutation relations with $y \in \Lambda_{o}$
\begin{equation}
 \{ c_{x, \sigma}, \alpha_{y}^{\dagger}(\zeta) \} =
 \lambda
 \{
 d_{x, \uparrow},
 a_{y, \uparrow}^{\dagger} +
 \zeta q^{y} a_{y, \downarrow}^{\dagger}
 \}
 = \lambda \eta_{x, \sigma} \delta_{x, y},
 \label{eq:anti-com_for_c_alpha-o}
\end{equation}
for $x \in \Lambda_{o}$, and
\begin{equation}
 \{ c_{x, \sigma}, \alpha_{y}^{\dagger}(\zeta)\}
 = - \eta_{x, \sigma}
 \left(
  q^{- \frac{1}{4}} \delta_{x - \frac{1}{2}, y}
  + q^{\frac{1}{4}} \delta_{x + \frac{1}{2}, y}
 \right)
 \label{eq:anti-com_for_c_alpha-prime}
\end{equation}
for $x \in \Lambda^{\prime}$ are useful to calculate the expectation
value, where $\eta_{x, \uparrow} = 1$ and
$\eta_{x, \downarrow} = \zeta q^{x}$.
These anticommutation relations (\ref{eq:anti-com_for_c_alpha-o}) and
(\ref{eq:anti-com_for_c_alpha-prime}) yield the following equation
\begin{equation}
 c_{x, \sigma} \Psi(\zeta) = {\rm sgn}[x] \eta_{x, \sigma}
 \Psi_{x}(\zeta),
\end{equation}
where $\Psi_{x}(\zeta)$ is defined by
\begin{equation}
 \Psi_{x}(\zeta) =
 \begin{cases}
  \displaystyle \lambda
  \left(
  \prod_{y \neq x}
  \alpha_{y}^{\dagger}(\zeta)
  \right)
  \Phi_{\rm vac} & \mbox{if } x \in \Lambda_{o} \\
  \displaystyle
  \left(
  q^{-\frac{1}{4}}
  \prod_{y \neq x - \frac{1}{2}}
  \alpha_{y}^{\dagger}(\zeta)
  - q^{\frac{1}{4}}
  \prod_{y \neq x + \frac{1}{2}}
  \alpha_{y}^{\dagger}(\zeta)
  \right)
  \Phi_{\rm vac}
  & \mbox{if } x \in \Lambda^{\prime}
 \end{cases}
 .
\end{equation}
The ordering manner of electron operators determines
${\rm sgn}[x] = \pm 1$, which is irrelevant to expectation values of any
local operators.
Then, the expectation value of $c_{x, \sigma}^{\dagger} c_{x, \tau}$ for
all $x \in \Lambda$ in the ground state $\Psi(\zeta)$ can be written in terms of
$\Psi_{x}(\zeta)$,
\begin{equation}
 \langle c_{x, \sigma}^{\dagger} c_{x, \tau} \rangle_{\zeta}
 =\eta_{x, \sigma}^{\ast} \eta_{x, \tau}
 \frac{\| \Psi_{x}(\zeta) \|^{2}}{\| \Psi(\zeta) \|^{2}}.
\end{equation}
Also a density one-point function $\langle n_{x} \rangle_{\zeta}$ can be
written in
\begin{equation}
 \langle n_{x} \rangle_{\zeta}
 =(1 + |\zeta q^{x}|^{2})
 \frac{\| \Psi_{x}(\zeta) \|^{2}}{\| \Psi(\zeta) \|^{2}},
\end{equation}
where $n_{x} = n_{x, \uparrow} + n_{x, \downarrow}$.
Thus we obtain the representations of spin one-point function at site
$x$ in terms of a density one-point function
$\langle n_{x} \rangle_{\zeta}$
\begin{equation}
 \left.
 \begin{array}{l}
  \displaystyle
  \langle S_{x}^{(1)} \rangle_{\zeta}
  = \frac{\langle n_{x} \rangle_{\zeta}}{2}
  \frac{\zeta q^{x} + (\zeta q^{x})^{\ast}}{1 + |\zeta q^{x}|^{2}},
  \vspace*{0.5em}\\
  \displaystyle
  \langle S_{x}^{(2)} \rangle_{\zeta}
  = \frac{\langle n_{x} \rangle_{\zeta}}{2 i}
  \frac{\zeta q^{x} - (\zeta q^{x})^{\ast}}{1 + |\zeta q^{x}|^{2}},
  \vspace*{0.5em}\\
  \displaystyle
  \langle S_{x}^{(3)} \rangle_{\zeta}
  = \frac{\langle n_{x} \rangle_{\zeta}}{2}
  \frac{1 - |\zeta q^{x}|^{2}}{1 + |\zeta q^{x}|^{2}},
 \end{array}
 \right\}
 .
 \label{eq:spin_function}
\end{equation}
Here we show briefly the estimates of the density one-point function.
We expect the density one-point function in the ground state
$\Psi(\zeta)$ takes almost a constant value in $\Lambda_{o}$ or in
$\Lambda^{\prime}$.
Indeed, we will prove this conjecture by evaluating the bounds of
this function. Here, we show only the asymptotic forms of the function
for large $|x - z|$. The details of evaluating the bounds are given in appendix
\ref{A:<n>}. The asymptotic forms of the density one-point function are
\begin{equation}
 \lim_{L \rightarrow \infty} \langle n_{x} \rangle_{\zeta}
  =
  \begin{cases}
   \displaystyle
   \frac{\lambda^{2}}{\sqrt{\epsilon^{2} - 4}}
   [1 + O(p^{-2|x - z|})]
   &\mbox{for} \quad x \in \Lambda_{o}\\
   \displaystyle
   \left(
   1 - \frac{\lambda^{2}}{\sqrt{\epsilon^{2} - 4}}
   \right)
   [1 + O(p^{-2|x - z|})]
   & \mbox{for} \quad x \in \Lambda^{\prime}
  \end{cases},
  \label{eq:n_asymp}
\end{equation}
where $\epsilon = \lambda^{2} + |q|^{\frac{1}{2}} + |q|^{-\frac{1}{2}}$
and $p = \min \{ r, |q| \}$ for $|q| > 1$ or $p = r$
for $|q| = 1$ with $r = (\epsilon + \sqrt{\epsilon^{2} - 4})/2$.
\begin{figure}[tb]
 \begin{center}
  \rotatebox{90}{$\langle n_{x} \rangle_{\zeta}$}%
  \begin{minipage}[c]{25pc}
   \begin{center}
    \setlength{\unitlength}{1pc}%
    \begin{picture}(0,0)(0,0)
     \put(-9,-6){(a)}
    \end{picture}
    \includegraphics[width=25pc]{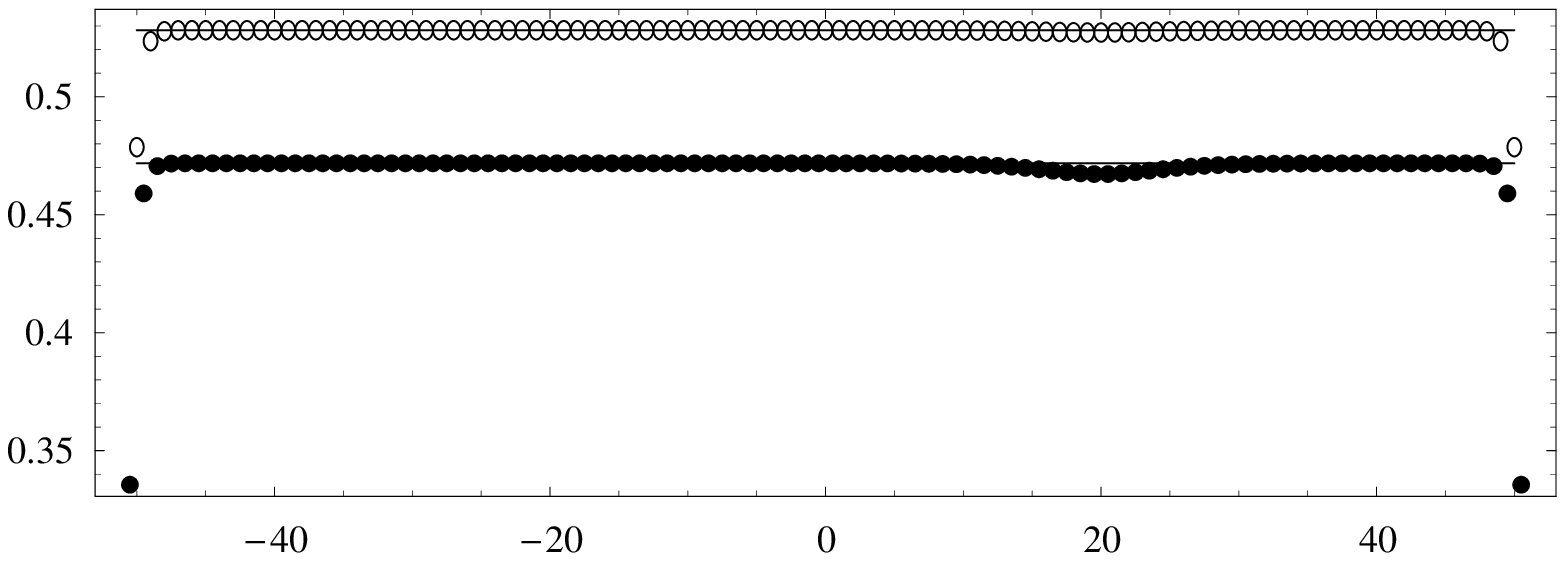}

    \hspace*{2em}$x$
   \end{center}
  \end{minipage}
  \vspace*{2em}

  \hspace*{-5mm}
  \rotatebox{90}{$\langle S_{x}^{(3)} \rangle_{\zeta}$}%
  \begin{minipage}[c]{25pc}
   \begin{center}
    \setlength{\unitlength}{1pc}%
    \begin{picture}(0,0)(0,0)
     \put(-9,-6){(b)}
    \end{picture}
    \includegraphics[width=25pc]{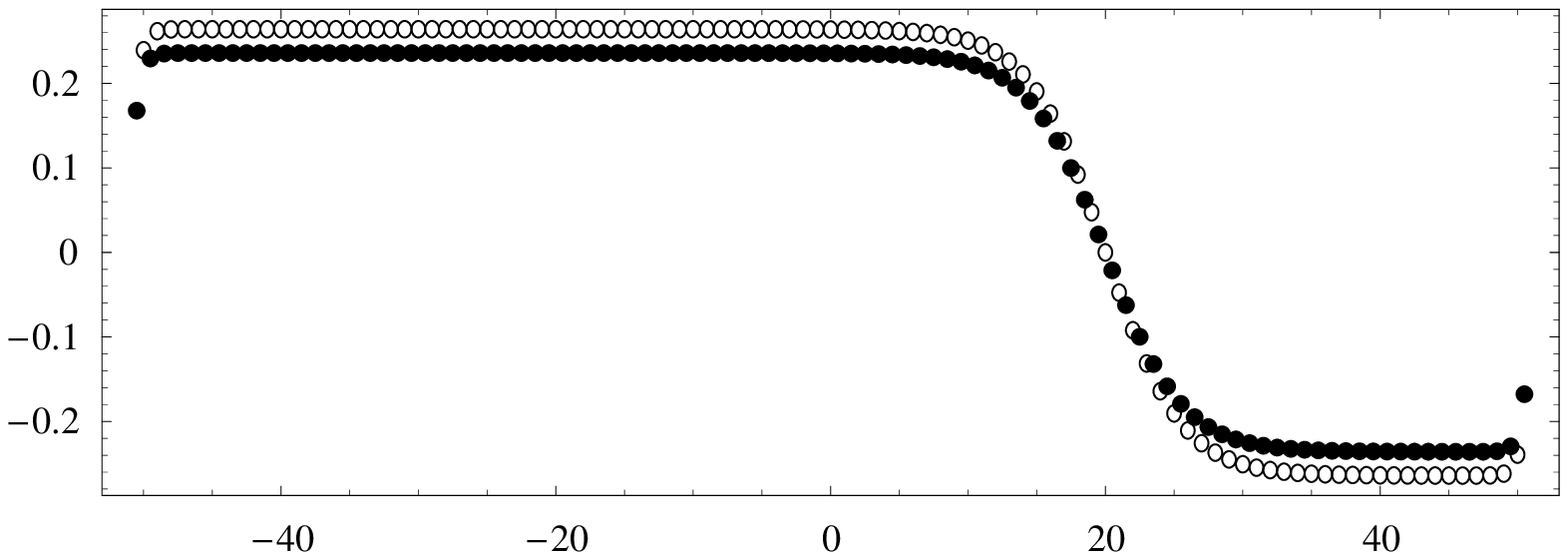}

    \hspace*{2em}$x$
   \end{center}
  \end{minipage}
 \end{center}
 \caption{The density one-point function (a) and the spin one-point
 function (b) are obtained by numerical calculation of rescaled
 normalization function for $L = 101$, $q = 1.2$, $\zeta = 1.2^{-20}$
 and $\lambda = 1.25$. The white circles plot the values on
 $\Lambda_{o}$ and the black dots plot the values on
 $\Lambda^{\prime}$. The lines in (a) are asymptotic values of density
 one-point function shown in Eq. (\ref{eq:n_asymp}) for $\Lambda_{o}$
 and $\Lambda^{\prime}$ respectively.}
 \label{fig:nd_and_s3}
\end{figure}
In these estimates, the density one-point function is shown to be
finite and strictly positive
at each site for any finite parameters.
Since the density one-point function is nearly equal to the asymptotic
value, the behavior of the spin one-point functions is not
controlled by the density one-point function besides the
oscillation between integer sites and half-odd-integer sites.
The spin one-point function $\langle S_{x}^{(3)} \rangle_{\zeta}$
vanishes only at $x=- \log_{|q|}|\zeta|$.
As discussed for the domain wall ground state in the XXZ models
\cite{KN1, KNS, Matsui2}, the two domains are distinguished by the sign of the
local order parameter $\langle S_{x} ^{(3)} \rangle_{\zeta}$.
If the domain wall center is defined by the zeros of
$\langle S_{x} ^{(3)} \rangle_{\zeta}$,
then the location of domain wall center is unique.
The function
$\frac{1}{2}\langle n_{x}\rangle_{\zeta}-|\langle S_{x}^{(3)}\rangle_{\zeta}|$
decays exponentially as the function of the distance from the center.
The domain wall width $1 / \log |q|$ is defined by its decay length.
In the large $\lambda$ limit in this model, the density one-point
function is exactly equal to 1 for integer sites and 0 for
half-odd-integer sites. In this limit and for real $q > 1$, electrons
are completely localized on integer sites, and the spin one-point
functions are exactly the same as those obtained in the XXZ model
defined on $\Lambda_{o}$.

For a complex $q = |q|e^{i \theta}$ with a real number $\theta$, one
can see the spiral structure with a pitch angle $\theta$. The vector
$\langle \vec S_x \rangle$ is rotated with the angle $\theta x$ around
the third spin axis depending on the site $x$.
Note that this spiral structure of the ground state does not exist in
the XXZ model.
Though the complex anisotropy parameter $q=e^{i \theta}$ is possible in
the XXZ Hamiltonian, the spiral state is no longer a ground state and
the corresponding model is described in the Tomonaga-Luttinger liquid
without ferromagnetic order. The translational symmetry in the
infinite-volume limit is broken by the domain wall or the spiral
structure for a finite $\log |\zeta|$. Both symmetries U(1) and
${\mathbb Z}_{2}$ are broken spontaneously as well.
We will discuss the broken symmetries in subsection
\ref{ssec:sym-in-TDL}.

\subsection{Electron Two-Point Function}

An electron two-point function
$\langle c_{x, \sigma}^{\dagger} c_{y, \tau} \rangle_{\zeta}$ can also
be written in terms of $\Psi_{x}(\zeta)$ and $\Psi_{y}(\zeta)$ by
\begin{align}
 \langle c_{x, \sigma}^{\dagger} c_{y, \tau} \rangle_{\zeta}
 =
 \frac{(c_{x, \sigma} \Psi(\zeta), c_{y, \tau} \Psi(\zeta))}
 {\| \Psi(\zeta) \|^{2}}
 =
 (-1)^{y - x} \eta_{x, \sigma}^{\ast} \eta_{y, \tau}
 \frac{(\Psi_{x}(\zeta), \Psi_{y}(\zeta))}{\| \Psi(\zeta) \|^{2}}.
\end{align}
We will see that the electron two-point function can be written in the
form:
\begin{equation}
 \langle c_{x, \sigma}^{\dagger} c_{y, \tau} \rangle_{\zeta}
 = K(x, y, L, \zeta)
 \frac{\eta_{x, \sigma}^{\ast} \eta_{y, \tau}}
 {\sqrt{1 + |\zeta q^{x}|^{2}} \sqrt{1 + |\zeta q^{y}|^{2}}}
 \left( -\frac{e^{i \frac{\theta}{2}}}{r} \right)^{|y - x|},
\end{equation}
where $K(x, y, L, \zeta)$ is a slowly varying finite function.
This expression shows that the electron two-point function decays
exponentially for large $|x-y|$ with the correlation length $1/\log r$.

\subsection{Spin Two-Point Functions}

The spin two-point functions can be written in the form of
\begin{equation}
 \langle S_{x}^{(j)} S_{y}^{(k)} \rangle_{\zeta}
 = \sum_{\sigma, \tau, \sigma^{\prime}, \tau^{\prime}}
 \frac{({\cal P}^{(j)})_{\sigma, \tau}}{2}
 \frac{({\cal P}^{(k)})_{\sigma^{\prime}, \tau^{\prime}}}{2}
 \langle
 c_{x, \sigma}^{\dagger} c_{x, \tau}
 c_{y, \sigma^{\prime}}^{\dagger} c_{y, \tau^{\prime}}
 \rangle_{\zeta}.
 \label{eq:def_of_SS}
\end{equation}
To obtain more explicit form, first, we rewrite the bracket into
\begin{equation}
 \langle c_{x, \sigma}^{\dagger} c_{x, \tau}
 c_{y, \sigma^{\prime}}^{\dagger} c_{y, \tau^{\prime}} \rangle_{\zeta}
 =
 \frac{
 (c_{x, \sigma} c_{y, \sigma^{\prime}} \Psi(\zeta),
 c_{x, \tau} c_{y, \tau^{\prime}} \Psi(\zeta))
 }{\| \Psi(\zeta) \|^{2}}.
\end{equation}
Using the anticommutator (\ref{eq:anti-com_for_c_alpha-o}) and
(\ref{eq:anti-com_for_c_alpha-prime}),
we can pick out the spin dependence as coefficients
\begin{equation}
 c_{x, \sigma} c_{y, \tau} \Psi(\zeta) =
 {\rm sgn}[x, y] \eta_{x, \sigma} \eta_{y, \tau} \Psi_{x, y}(\zeta),
\end{equation}
where ${\rm sgn}[x,y] = \pm 1$ and
\begin{equation}
 \Psi_{x, y}(\zeta) =
 \begin{cases}
 \displaystyle
 \lambda^{2}
 \left[ \prod_{w \neq x, y} \alpha_{w}^{\dagger}(\zeta) \right]
 \Phi_{\rm vac}
 & \mbox{if } x, y \in \Lambda_{o}\\
 \displaystyle
 \lambda
 \left[
 q^{-\frac{1}{4}}
 \prod_{w \neq x, y - \frac{1}{2}} \alpha_{w}(\zeta)^{\dagger}
 -
 q^{\frac{1}{4}}
 \prod_{w \neq x, y + \frac{1}{2}} \alpha_{w}(\zeta)^{\dagger}
 \right] \Phi_{\rm vac}
 & \mbox{if } x \in \Lambda_{o}, y \in \Lambda^{\prime} \\
 \displaystyle
 \lambda
 \left[
 q^{-\frac{1}{4}}
 \prod_{w \neq x - \frac{1}{2}, y} \alpha_{w}(\zeta)^{\dagger}
 -
 q^{\frac{1}{4}}
 \prod_{w \neq x + \frac{1}{2}, y} \alpha_{w}(\zeta)^{\dagger}
 \right] \Phi_{\rm vac}
 & \mbox{if } y \in \Lambda_{o}, x \in \Lambda^{\prime}\\
 \displaystyle
 \left[
 q^{-\frac{1}{2}}
 \prod_{w \neq x - \frac{1}{2}, y - \frac{1}{2}}
 \alpha_{w}(\zeta)^{\dagger}
 +
 q^{\frac{1}{2}}
 \prod_{w \neq x + \frac{1}{2}, y + \frac{1}{2}}
 \alpha_{w}(\zeta)^{\dagger}
 \right. & \\
 \displaystyle
 \quad \left.
 -
 \prod_{w \neq x + \frac{1}{2}, y - \frac{1}{2}}
 \alpha_{w}(\zeta)^{\dagger}
 -
 \prod_{w \neq x - \frac{1}{2}, y + \frac{1}{2}}
 \alpha_{w}(\zeta)^{\dagger}
 \right] \Phi_{\rm vac}
 & \mbox{if } x, y \in \Lambda^{\prime}\\
 \end{cases}
 .
\end{equation}
Then the bracket in (\ref{eq:def_of_SS}) is reduced to
\begin{equation}
 \langle c_{x, \sigma}^{\dagger} c_{x, \tau}
 c_{y, \sigma^{\prime}}^{\dagger} c_{y, \tau^{\prime}} \rangle_{\zeta}
 =
 \eta_{x, \sigma}^{\ast} \eta_{x, \tau}
 \eta_{y, \sigma^{\prime}}^{\ast} \eta_{y, \tau^{\prime}}
 \frac{\| \Psi_{x, y}(\zeta) \|^{2}}{\| \Psi(\zeta) \|^{2}}.
\end{equation}
Note that the density two-point function is written
\begin{equation}
 \langle n_{x} n_{y} \rangle_{\zeta}
 = (1 + |\zeta q^{x}|^{2})(1 + |\zeta q^{y}|^{2})
 \frac{\| \Psi_{x, y}(\zeta) \|^{2}}{\| \Psi(\zeta) \|^{2}}.
\end{equation}
We then obtained the representation of the spin two-point
function in terms of the density two-point function as well as
the spin one-point function written in terms of the density one-point
function
\begin{align}
 \langle S_{x}^{(j)} S_{y}^{(k)} \rangle_{\zeta}
 = &
 \sum_{\sigma, \tau, \sigma^{\prime}, \tau^{\prime}}
 \frac{
 \eta_{x, \sigma}^{\ast}
 \frac{1}{2}({\cal P}^{(j)})_{\sigma, \tau} \eta_{x, \tau}
 }
 {1 + |\zeta q^{x}|^{2}}
 \frac{
 \eta_{y, \sigma^{\prime}}^{\ast}
 \frac{1}{2}{(\cal P}^{(k)})_{\sigma^{\prime}, \tau^{\prime}}
 \eta_{y, \tau^{\prime}}
 }{1 + |\zeta q^{y}|^{2}}
 \langle n_{x} n_{y} \rangle_{\zeta}.
\end{align}
If $\lambda$ is finite, we can rewrite spin two-point function from the
spin one-point function (\ref{eq:spin_function})
\begin{equation}
 \langle S_{x}^{(j)} S_{y}^{(k)} \rangle_{\zeta}
 = \langle S_{x}^{(j)} \rangle_{\zeta}
 \langle S_{y}^{(k)} \rangle_{\zeta}
 \frac{\langle n_{x} n_{y} \rangle_{\zeta}}
 {\langle n_{x} \rangle_{\zeta} \langle n_{y} \rangle_{\zeta}}.
\end{equation}
In section \ref{sec:infinite}, we will find that
$\langle n_{x} \rangle_{\zeta}$ and
$\langle n_{x} n_{y} \rangle_{\zeta}$ exist in the limit
$|x| \rightarrow \infty$, $|y| \rightarrow \infty$ and both.
And we will prove
\begin{equation}
 \lim_{|x - y| \rightarrow \infty} \lim_{L \rightarrow \infty}
  \bigl(
  \langle n_{x} n_{y} \rangle_{\zeta}
  - \langle n_{x} \rangle_{\zeta} \langle n_{y} \rangle_{\zeta}
  \bigr) = 0,
\end{equation}
namely,
\begin{equation}
 \lim_{|x-y| \rightarrow \infty} \lim_{L \rightarrow \infty}
 \frac{ \langle n_{x} n_{y} \rangle_{\zeta}}
 {\langle n_{x} \rangle_{\zeta} \langle n_{y} \rangle_{\zeta}}
 =1.
\end{equation}
Thus, we conclude that the spin two-point function satisfies the
requirement of the cluster property.
We can also find bounds of the density
two-point function. The function behaves asymptotically
\begin{equation}
 \lim_{L \rightarrow \infty}
  |\langle n_{x} n_{y} \rangle_{\zeta}
  - \langle n_{x} \rangle_{\zeta} \langle n_{y} \rangle_{\zeta}|
  =
  \begin{cases}
   O(p^{-2|x - z|})
   & \mbox{if } |y - x| > |x - z| > |y - z| \\
   O(p^{-2|y - z|})
   & \mbox{if } |y - x| > |y - z| > |x - z| \\
   O(p^{-2|y - x|})
   & \mbox{otherwise}
  \end{cases}
\end{equation}
where $p$ has the same value as that appeared in Eq.(\ref{eq:n_asymp}).
We obtain these decay relations in appendix \ref{A:tran-corr-func}.
The spin two-point function is almost a product of two
spin one-point functions.
This fact shows that the electron has a good localization property. In
other words, $\Psi(\zeta)$ is almost a product state.
In the large $\lambda$ limit,
$\langle n_{x} n_{y} \rangle_{\zeta}- \langle n_{x} \rangle_{\zeta} \langle n_{y} \rangle_{\zeta}$
becomes $0$ identically, and then the spin two-point function becomes
just a product of two spin one-point functions.
This property is the same as domain wall ground state in the XXZ
model. We believe that our model is related to the XXZ model if
$q \in {\mathbb R}$.

\section{Normalization Function \label{sec:N-func}}

In this section, we introduce a normalization function which is useful
to evaluate the density one-point function and two-point
functions in the ground state $\Psi(\zeta)$.

\subsection{Definition of Normalization Function}

We define a normalization function $A(x, y; \zeta)$ for
$x, y \in \Lambda_{o}$ with $x \leq y$ by
\begin{equation}
 A(x, y; \zeta) =
 \left\|
  \left(
  \prod_{z = x}^{y} \alpha_{z}^{\dagger}(\zeta)
  \right)
  \Phi_{\rm vac}
 \right\|^{2}.
 \label{norm}
\end{equation}
Note that the norm of the ground state $\Psi(\zeta)$ can be represented by
$\| \Psi(\zeta) \|^{2}=A(-l, l; \zeta)$, where $l = (L - 1)/2$.
The anticommutation relation for $\alpha_{x}(\zeta)$ and
$\alpha_{y}^{\dagger}(\zeta)$ for $x, y \in \Lambda_{o}$ is
\begin{equation}
 \{ \alpha_{x}(\zeta), \alpha_{y}^{\dagger}(\zeta) \}
 = ( 1 + |\zeta q^{x - \frac{1}{2}}|^{2}) e^{-i \frac{\theta}{2}}
 \delta_{x - 1, y}
 + \epsilon (1 + |\zeta q^{x}|^{2}) \delta_{x, y}
 + (1 + |\zeta q^{x + \frac{1}{2}}|^{2}) e^{i \frac{\theta}{2}}
 \delta_{x + 1, y}
 \label{eq:AC_of_alpha's}
\end{equation}
where
$\epsilon = \lambda^{2} + |q|^{\frac{1}{2}} + |q|^{-\frac{1}{2}}$.
Using this anticommutation relation, we can obtain the recursion
relation for $A(x, y; \zeta)$ in terms of $A(x, y - 1; \zeta)$ and
$A(x, y - 2; \zeta)$, or $A(x + 1, y; \zeta)$ and $A(x + 2, y; \zeta)$:
\begin{equation}
 A(x, y; z) =
 \epsilon (1 + |\zeta q^{y}|^{2}) A(x, y - 1; \zeta)
 -(1 + |\zeta q^{y - \frac{1}{2}}|^{2})^{2} A(x, y - 2; \zeta),
 \label{eq:rec_of_A1}
\end{equation}
or
\begin{equation}
 A(x, y; z) =
 \epsilon (1 + |\zeta q^{x}|^{2}) A(x + 1, y; \zeta)
 -(1 + |\zeta q^{x + \frac{1}{2}}|^{2})^{2} A(x + 2, y; \zeta).
 \label{eq:rec_of_A2}
\end{equation}
This forms show that we can obtain $A(x, y; \zeta)$ recursively from the
boundary values $A(x, x; \zeta)$ and $A(x, x + 1; \zeta)$, or
$A(y, y; \zeta)$ and $A(y - 1, y; \zeta)$, which are easily calculated
as
\begin{equation}
 \left.
 \begin{array}[c]{l@{=}l}
  \displaystyle
  A(x, x; \zeta) &
  \displaystyle
  \epsilon (1 + |\zeta q^{x}|^{2}) \vspace*{1mm}\\
  \displaystyle
  A(x, x + 1; \zeta) &
  \displaystyle
  \epsilon^{2} (1 + |\zeta q^{x}|^{2}) (1 + |\zeta q^{x + 1}|^{2})
  -(1 + |\zeta q^{x + \frac{1}{2}}|^{2})^{2} \vspace*{1mm}\\
  \displaystyle
  A(y - 1, y; \zeta) &
  \displaystyle
  \epsilon^{2} (1 + |\zeta q^{y - 1}|^{2}) (1 + |\zeta q^{y}|^{2})
  -(1 + |\zeta q^{y - \frac{1}{2}}|^{2})^{2}
 \end{array}
 \right\}.
 \label{eq:bb_of_A}
\end{equation}
The following formal boundary values are convenient
\begin{equation}
 A(x, x - 1; \zeta) = A(x + 1, x; \zeta) = 1,~~
 A(x, x - 2; \zeta) = A(x + 2, x; \zeta) = 0.
 \label{eq:def-A+-}
\end{equation}
We can also represent
$A(x, y; \zeta)$ in terms of the values $A(w + \delta , y; \zeta)$ and
$A(x, w - \delta; \zeta)$ for $w \in \Lambda_{o}$ and $\delta = 1, 2$
with $x \leq w \leq y$,
\begin{align}
 A(x, y; \zeta) =&
 - (1 + |\zeta q^{w - \frac{1}{2}}|^{2})^{2}
 A(x, w - 2; \zeta) A(w + 1, y; \zeta)
 \nonumber \\
 & + \epsilon (1 + |\zeta q^{w}|^{2})
 A(x, w - 1; \zeta) A(w + 1, y; \zeta)
 \nonumber \\
 & - (1 + |\zeta q^{w + \frac{1}{2}}|^{2})^{2}
 A(x, w - 1; \zeta) A(w + 2, y; \zeta), \label{eq:dev_of_A}
\end{align}
where we can use the formally defined boundary values
(\ref{eq:def-A+-}). Note that the relations (\ref{eq:rec_of_A1}),
(\ref{eq:rec_of_A2}) and (\ref{eq:dev_of_A}) are equivalent to each other
with the formal boundary values (\ref{eq:def-A+-}).
Hereafter, we consider that $A(x, y; \zeta)$ is a function of
$x, y \in {\mathbb Z}$ and $\zeta \in {\mathbb C}$ with $y - x \geq -2$,
which is defined by the recursion relation (\ref{eq:rec_of_A1}) and formal
boundary values (\ref{eq:def-A+-}).

If $|q| = 1$, then the coefficients of the recursion relation
(\ref{eq:rec_of_A1}) are independent of $y$.
This condition gives a great advantage for calculations. We define
$\widetilde{A}(x, y)$ by
\begin{equation}
 A(x, y; \zeta) = (1 + |\zeta|^{2})^{y - x + 1}
 \widetilde{A}(x, y).
\end{equation}
The recursion relation and boundary values for $\widetilde{A}(x, y)$ are
obtained from those for the original function $A(x, y; \zeta)$,
\begin{equation}
 \widetilde{A}(x, y) =
 \epsilon \widetilde{A}(x, y - 1)
 - \widetilde{A}(x, y - 2),
\end{equation}
and
\begin{equation}
 \widetilde{A}(x, x - 2) = \widetilde{A}(y + 2, y) = 0,
 \quad
 \widetilde{A}(x, x - 1) = \widetilde{A}(y + 1, y) = 1.
\end{equation}
By rewriting this recursion relation to
\begin{equation}
 \widetilde{A}(x, y) - r \widetilde{A}(x, y - 1)
 = r^{-1}
 \left[
  \widetilde{A}(x, y - 1) - r \widetilde{A}(x, y - 2)
 \right],
\end{equation}
where $r = (\epsilon + \sqrt{\epsilon^{2} - 4})/2$, we can easily obtain
the solution
\begin{equation}
 \widetilde{A}(x, y) = \frac{r^{y - x + 2} - r^{-(y - x + 2)}}{r - r^{-1}}.
\end{equation}
Then we obtain a compact representation of the function
\begin{equation}
 A(x, y; \zeta) =
 \left[
  r (1 + |\zeta|^{2})
 \right]^{y - x + 1}
 \frac{1 - r^{-2 (y - x + 2)}}{1 - r^{-2}}.
 \label{eq:A_with_|q|=1}
\end{equation}

\subsection{Rescaled Normalization Function}

In the case of $|q| \neq 1$, we parametrize $|\zeta|$ by a real number $z$
as $|\zeta| = |q|^{- z}$ and define a rescaled normalization function
$B$ by
\begin{equation}
 A(x, y; \zeta) =
 \left[
  \prod_{w = x}^{y} r \left( 1 + |q|^{2(w - z)} \right)
 \right] B(x, y, z).
 \label{normres}
\end{equation}
The corresponding recursion relation for the rescaled function
obtained from those for the original function (\ref{eq:rec_of_A1}) is
written 
\begin{equation}
 B(x, y, z) =
  \displaystyle
   \frac{1 + r^{2}}{r^{2}} B(x, y - 1, z)
   -\frac{1 - f(y - z - \frac{1}{2})}{r^{2}} B(x, y - 2, z)
\label{eq:rec_of_B}
\end{equation}
where $f$ is defined by
\begin{equation}
 f(u) =
 \frac{(|q|^{\frac{1}{2}} - |q|^{-\frac{1}{2}})^{2}}
 {(|q|^{- u + \frac{1}{2}} + |q|^{u - \frac{1}{2}})
 (|q|^{- u - \frac{1}{2}} + |q|^{u + \frac{1}{2}})}.
\end{equation}
This $f(u)$ is an even function of a real number $u$. At $u = 0$,
$f(u)$ takes the maximal value
\begin{equation}
 f(0) =
 \left(
  \frac{|q|^{\frac{1}{2}} - |q|^{-\frac{1}{2}}}
  {|q|^{\frac{1}{2}} + |q|^{-\frac{1}{2}}}
 \right)^{2},
\end{equation}
and for large $|u|$, $f(u)$ is proportional to $|q|^{-2|u|}$
asymptotically.
The boundary values are
\begin{equation}
 B(x, x - 1, z) = B(x + 1, x, z) = 1,~~
 B(x, x - 2, z) = B(x + 2, x, z) = 0. \label{eq:bb_of_B}
\end{equation}
The relations obtained from (\ref{eq:rec_of_A2}) and
(\ref{eq:dev_of_A}) are
\begin{equation}
 B(x, y, z) =
 \frac{1 + r^{2}}{r^{2}} B(x + 1, y, z)
 -\frac{1 - f(x - z + \frac{1}{2})}{r^{2}} B(x + 2, y, z)
 \label{eq:rec_of_B2}
\end{equation}
and
\begin{align}
 B(x, y, z) =
 & \displaystyle
 -\frac{1 - f(w - \frac{1}{2} - z)}{r^{2}}
 B(x, w - 2, z) B(w + 1, y, z)
 \nonumber \\
 &
 \displaystyle
 + \frac{1 + r^{2}}{r^{2}} B(x, w - 1, z) B(w + 1, y, z)
 \nonumber \\
 &
 \displaystyle
 -\frac{1 - f(w + \frac{1}{2} - z)}{r^{2}}
 B(x, w - 1, z) B(w + 2, y, z),
 \label{eq:dev_of_B}
\end{align}
which are equivalent to the recursion relation (\ref{eq:rec_of_B}).

Let us discuss properties of the rescaled function.
The function $B(x, y, z)$ is non-negative and monotonically
increasing in $y - x$, i.e.
\begin{equation}
 B(x, y, z) \geq B(x, y - 1, z),~~
 B(x, y, z) \geq B(x + 1, y, z).
\end{equation}
To obtain these relations, we rewrite the recursion relation
(\ref{eq:rec_of_B})
\begin{equation}
 B(x, y, z) - B(x, y - 1, z) =
 \frac{B(x, y - 1, z) - B(x, y - 2, z)}{r^{2}}
 +\frac{f(y - \frac{1}{2} - z)}{r^{2}}B(x, y - 2, z).
\end{equation}
This representation shows $B(x, y, z) \geq B(x, y - 1, z)$ if
$B(x, y - 1, z) \geq B(x, y - 2, z)$ and $B(x, y - 2 , z) \geq 0$.
From the
formal definition (\ref{eq:bb_of_B}),
we know $B(x, x - 1, z) > B(x, x - 2) = 0$ , and therefore
$B(x, y, z) \geq B(x, y - 1, z) > 0$.
Also $B(x, y, z) \geq B(x + 1, y, z)$ is shown almost in the same
way.
The function $B(x, y, z)$ is invariant under replacing $|q|$ by
$|q|^{-1}$ since the coefficients in recursion relation
(\ref{eq:rec_of_B}) and the formal boundary values are invariant under the 
replacement of $|q|$ by $|q|^{-1}$.

Though we cannot obtain a compact form of the solution $B(x, y, z)$, we
show that $B(x, y, z)$ exists in each limit $y \rightarrow \infty$,
$x \rightarrow - \infty$ or both respectively in the next subsection.
And we can obtain bounds of the ratio of rescaled normalization
functions at two different sites. This bounds
are useful to estimate the expectation values in the
ground state $\Psi(\zeta)$. We will obtain these bounds in appendix
\ref{A:RNF}.

\subsection{Convergence of the Rescaled Function}

Let us prove that $B(x, y, z)$ converges in each limit
$y \rightarrow \infty$, $x \rightarrow - \infty$ or both.
We use the relation (\ref{eq:dev_of_B}) for these estimates.
The existence of $B(-\infty, \infty, z)$ is shown by this relation,
if we show the existence of $B(x, \infty, z)$ and $B(-\infty, y, z)$.
We then concentrate to show the convergence 
in the limit $y \rightarrow \infty$.
The convergence in another limit
$x \rightarrow - \infty$ can be shown in the same way.

We introduce new functions $C(x, y, z)$ and $D(x, y)$ where
$x, y \in {\mathbb Z}$ and $z \in {\mathbb R}$ with $y - x \leq -2$.
These functions are defined by following recursion relations:
\begin{equation}
 D(x, y) =
 \frac{1 + r^{2}}{r^{2}} D(x, y - 1)
 -\frac{1 - g}{r^{2}} D(x, y - 2)
 \label{eq:rec_of_D},
\end{equation}
and
\begin{align}
 C(x, y, z) =
 \frac{1 + r^{2}}{r^{2}}
 C(x, y - 1, z)
 - \frac{1}{r^{2}} C(x, y - 2, z)&
 \nonumber \\
 + \frac{f(y - \frac{1}{2} - z)}{r^{2}}
  D(x, y - 2)&.
  \label{eq:rec_of_C}
\end{align}
with initial values
\begin{equation}
 C(x, x - 2, z) = D(x, x - 2) = 0, ~~
 C(x, x - 1, z) = D(x, x - 1) = 1,
\end{equation}
where $g$ is a real parameter.
Note that both functions $C(x, y, z)$ and $D(x, y)$ are
nonnegative and monotonically increasing for $y - x$ as well as the rescaled
normalization function $B(x, y, z)$.
Both recursion relations for $C(x, y, z)$ and $D(x, y)$ are much easier than that for $B(x,y,z)$.
One can easily obtain a compact representation of the solution $D(x, y)$
\begin{equation}
 D(x, y) =
  \frac{R_{+}^{y - x + 2} - R_{-}^{y - x + 2}}
 {R_{+} - R_{-}} \label{eq:sol_of_D}
\end{equation}
where $R_{\pm}$ are defined by
\begin{equation}
 R_{+} = 1 + \gamma (g) ~\mbox{ and }~
 R_{-} = r^{-2} - \gamma (g)
\end{equation}
with
\begin{equation}
 \gamma (g) =
 \frac{\sqrt{\epsilon^{2} - 4 + 4 g} - \sqrt{\epsilon^{2} - 4}}{2 r}.
\end{equation}
Note that $\gamma(g)$ takes a range $0 < \gamma(g) < r^{-2}$ for
$0 < g < 1$.

If $g$ satisfies
\begin{equation}
 g \geq \max_{x \leq w \leq y - 1} f({\textstyle w + \frac{1}{2} - z}),
 \label{eq:cond_for_g}
\end{equation}
these functions satisfy the following inequalities,
\begin{equation}
 B(x, y, z) \leq C(x, y, z) \leq D(x, y),
 \label{eq:inequ_for_BCD}
\end{equation}
and
\begin{equation}
 B(x, y, z) - B(x, y - 1, z) \leq
 C(x, y, z) - C(x, y - 1, z), \label{eq:inequ_for_B_diff}
\end{equation}
for any integers $x, y$ with $x \leq y$ and a real number $z$.
To show the second inequality in (\ref{eq:inequ_for_BCD}), we
calculate $D(x, y) - C(x, y, z)$ using the recursion relation
for each
\begin{align}
 & D(x, y) - C(x, y, z)
 - [D(x, y - 1) - C(x, y - 1, z)]
 \nonumber \\
 & =
 \frac{1}{r^{2}}
 \bigl\{
 D(x, y - 1) - C(x, y - 1, z)
 - [D(x, y - 2) - C(x, y - 2, z)]
 \bigr\}
 \nonumber \\
 & \quad
 + \frac{f(y - \frac{1}{2} - z)}{r^{2}}
 \left[D(x, y - 2) - C(x, y - 2, z) \right]
 + \frac{g - f(y - \frac{1}{2} - z)}{r^{2}} D(x, y - 2).
 \label{eq:D-C}
\end{align}
And note
\begin{equation}
 D(x, x) - C(x, x, z) = 0 ~\mbox{and}~
 D(x, x + 1) - C(x, x + 1, z) =
 \frac{g - f(x + \frac{1}{2} - z)}{r^{2}} > 0.
 \label{eq:D-C_initial}
\end{equation}
Eqs. (\ref{eq:D-C}), (\ref{eq:D-C_initial}) and the condition
(\ref{eq:cond_for_g}) imply that $D(x, y) - C(x, y, z)$ is non-negative
and monotonically increasing function of $y - x > 0$ because $D(x, y) > 0$.
Thus, we obtain $D(x, y) \geq C(x, y, z)$.
The remaining two inequalities in (\ref{eq:inequ_for_BCD}) and
(\ref{eq:inequ_for_B_diff}) can be shown in the same way.

Now we prove the convergence of $C(x, y, z)$ for $g = f(0)$ in the limit
of $y \rightarrow \infty$. The convergence of $C(x, y, z)$ 
and the first inequality in (\ref{eq:inequ_for_BCD})
give the convergence of $B(x, y, z)$.
We rewrite the recursion relation for $C(x, y, z)$
\begin{align}
 C(x, y, z) - C(x, y - 1, z)
 =
 \frac{C(x, y - 1, z) - C(x, y - 2, z)}{r^{2}}
 \nonumber \\
 + \frac{f(y - \frac{1}{2} - z)}{r^{2}}
 \frac{R_{+}^{y - x} - R_{-}^{y - x}}{R_{+} - R_{-}}.
 \label{eq:rec_of_C2}
\end{align}
The final term can be evaluated by
\begin{align}
 & \frac{f(y - \frac{1}{2} - z)}{r^{2}}
 \frac{R_{+}^{y - x} - R_{-}^{y - x}}{R_{+} - R_{-}}
 \nonumber \\
 & \qquad =
 \left( \frac{R_{+}}{|q|^{2}} \right)^{y}
 \frac{1}{r^{2}}
 \frac{
 (|q|^{\frac{1}{2}} - |q|^{-\frac{1}{2}})^{2}
 }
 {
 |q|^{-2 z - 1} + |q|^{-4 y + 2 z + 1}
 + |q|^{-2 y - 1} + |q|^{-2 y + 1}
 }
 \frac{R_{+}^{-x} - R_{-}^{y - x}/R_{+}^{y}}
 {R_{+} - R_{-}}
 \nonumber \\
 & \qquad <
 \left( \frac{R_{+}}{|q|^{2}} \right)^{y}
 \frac{|q|^{2 z + 1}}{r^{2} R_{+}^{x}}
 \frac{
 (|q|^{\frac{1}{2}} - |q|^{-\frac{1}{2}})^{2}}
 {R_{+} - R_{-}}.
 \nonumber
\end{align}
From the recursion (\ref{eq:rec_of_C2}), we have
\begin{equation}
 C(x, y, z) - C(x, y - 1, z)
 \leq
 \frac{C(x, y - 1, z) - C(x, y - 2, z)}{r^{2}}
 +
 \beta^{y}
 K_{1}(x, z) \label{eq:inequ_for_C-C-pre}
\end{equation}
where $\beta = R_{+}/|q|^{2}$ and
\begin{equation}
 K_{1}(x, z)
 =
 \frac{|q|^{2 z + 1}}{r^{2} R_{+}^{x}}
 \frac{
 (|q|^{\frac{1}{2}} - |q|^{-\frac{1}{2}})^{2}}
 {R_{+} - R_{-}}.
\end{equation}
Here, we show $\beta < 1$ for $|q| > 1$ and $g \leq f(0)$.
We define a function
\[
 h(\epsilon, g) \equiv
 \frac{\epsilon + \sqrt{\epsilon^{2} - 4 + 4 g}}
 {\epsilon + \sqrt{\epsilon^{2} - 4}},
\]
which is monotonically decreasing in $\epsilon \geq 2$ and monotonically
increasing in $g \geq 0$. Then,
we represent $\beta$ as
\begin{equation}
\beta = \frac{1}{|q|^{2}} h(\epsilon, g).
\label{eq:beta}
\end{equation}
Since $\epsilon = \lambda^2 + |q|^{\frac{1}{2}} +|q|^{-\frac{1}{2}} \geq 2$, $g \leq f(0)$ and $|q| > 1$,
we obtain
\begin{equation}
\beta = \frac{1}{|q|^{2}} h(\epsilon,g ) \leq \frac{1}{|q|^{2}} h(2, f(0))=
\frac{2}{|q|^{\frac{3}{2}} (|q|^{\frac{1}{2}} + |q|^{-\frac{1}{2}})}
  < 1.
 \label{eq:lowest_of_beta}
\end{equation}
Using (\ref{eq:inequ_for_C-C-pre}) recursively, we obtain
\begin{align}
 C(x, y, z) - C(x, y - 1, z)
 \leq &
 \frac{C(x, y - 2, z) - C(x, y - 3, z)}{r^{4}}
 + \frac{\beta^{y - 1}}{r^{2}} K_{1}(x, z)
 + \beta^{y} K_{1}(x, z)
 \nonumber \\
 \vdots &
 \nonumber \\
 \leq &
 \frac{1}{r^{2(y - x + 1)}}
 +
 \beta^{y}
 \frac{(r^{2} \beta)^{- (y - x + 1)} - 1}{(r^{2} \beta)^{-1} - 1}
 K_{1}(x, z)
 \nonumber \\
 = & K_{2}(z - x) r^{-2 (y - x)} + K_{3}(z - x) \beta^{y - x}
\end{align}
where
\begin{equation}
 K_{2}(z - x) =
 \frac{1}{r^{2}}
 \left[
 1 +
 \frac{|q|^{2 (z - x) + 3}}{|q|^{2} - r^{2} R_{+}}
 \frac{
 (|q|^{\frac{1}{2}} - |q|^{-\frac{1}{2}})^{2}}
 {R_{+} - R_{-}}
 \right]
\end{equation}
and
\begin{equation}
 K_{3}(z - x) =
 \frac{R_{+} |q|^{2 (z - x) + 1}}{r^{2} R_{+} - |q|^{2}}
 \frac{
 (|q|^{\frac{1}{2}} - |q|^{-\frac{1}{2}})^{2}}
 {R_{+} - R_{-}}.
\end{equation}
$K_{2, 3}(z - x)$ are finite for fixed $x$ and $z$.
Then we have found that the difference between $C(x, y, z)$ and
$C(x, y - 1, z)$ decays exponentially as $y$ becomes large if $x$ and
$z$ are fixed, and hence $C(x, y, z)$ converges for
$y \rightarrow \infty$.
This fact means that the limit $B(x, \infty, z)$ exists. 
As discussed before, we easily find
the convergence in the limit of $x \rightarrow -\infty$ and in the both
$y \rightarrow \infty$ and $x \rightarrow -\infty$.

\section{The Infinite-Volume Limit \label{sec:infinite}}

In this section, we consider the infinite-volume limit.
First, we define a ground state $\omega_\zeta$ in the infinite-volume
system on the basis of the ground state $\Psi(\zeta)$ in the
finite-volume system. We show that any local operator cannot deform the
ground state $\omega_\zeta$ to a state with lower energy than that of
$\omega_\zeta$. Next, we remark the symmetries of the model Hamiltonian
in the infinite-volume limit and discuss their breakdown.
We show also the cluster property of
the ground state $\omega_\zeta$ and excited states above $\omega_\zeta$.
For the general proof, we use the representations in terms of the
rescaled normalization function which is defined only for
$|q| \neq 1$ in this paper. For $|q| = 1$, however, the proof is
much easier.

\subsection{Ground State in the Infinite-Volume System}

A state $\omega_{\zeta}$ is a linear functional
over all local operators for the infinite-volume system defined by
\begin{equation}
 \omega_{\zeta} ( O_{\A} ) := \lim_{L \rightarrow \infty}
  \langle O_{\A} \rangle_{\zeta}
= \lim_{L \rightarrow \infty}
  \frac{( \Psi(\zeta), O_{\A} \Psi(\zeta))}{\| \Psi(\zeta) \|^{2}},
\end{equation}
for an arbitrary local operator $O_{\A}$ written in electron operators
on a finite subset $\A$ of the lattice $\Lambda$. We define also other
states $\omega_{\uparrow}$ and $\omega_{\downarrow}$ for the
infinite-volume system corresponding to the finite-volume ground states
$\Phi_{\uparrow}$ and $\Phi_{\downarrow}$ with the definition
(\ref{allup}).

This state $\omega_{\zeta}$ is the ground state in the infinite-volume
system, if the inequality
\begin{equation}
\lim_{L_1 \rightarrow \infty}  \omega_{\zeta} ( O_{\A}^{\dagger}
[H_{L_1}, O_{\A}])
  \geq 0,
  \label{eq:LS}
\end{equation}
is valid for any local operator $ O_{\A}$,
where $H_{L_1}$ is a local Hamiltonian defined on a finite chain with
length $L_1$. This inequality implies that any local operator cannot decrease the
energy, and hence we can regard the state  $\omega_{\zeta}$ as a ground state
in the infinite-volume limit.
In the Hamiltonian (\ref{eq:H_tot}), there exists
$L_{2} \in {\mathbb Z}$ for the set $\A$ such that
\begin{equation}
\lim_{L_1 \rightarrow \infty}  [O_{\A}, H_{L_1}] = [O_{\A}, H_{L_{2}}].
\end{equation}
Since $H_{L_{2}}$ is positive semi-definite and
$H_{L_{2}} \Psi(\zeta) = 0$ for any finite $L_2$,
\begin{equation}
 \frac{( \Psi(\zeta), O_{\A}^{\dagger} [H_{L_{_2}}, O_{\A}] \Psi(\zeta))}
  {\| \Psi(\zeta) \|^{2}} \geq 0
\end{equation}
is trivial. Thus,
$\omega_{\zeta}$ satisfies the inequality (\ref{eq:LS}).
We can easily verify that the states $\omega_{\uparrow}$ and $\omega_{\downarrow}$
are the ground states of the infinite-volume system.

\subsection{Symmetries and Their Breakdown\label{ssec:sym-in-TDL}}

Here, we consider the U(1), the ${\mathbb Z}_2$ and the translational
symmetries of the Hamiltonian in the infinite-volume system. These
symmetries are defined for a finite system by Eqs.(\ref{U}),
(\ref{eq:pi}) and (\ref{T}). Let $\G$ be one symmetry group of them. An
arbitrary transformation $X \in \G$ can act on an arbitrary local
operator $O_{\A}$ written in electron operators 
on a finite subset $\A$ of the lattice $\Lambda$
\[
 O_{\A} \longmapsto X(O_{\A} ).
\]
The Hamiltonian of the infinite-volume system has a
symmetry $\G$, if a local Hamiltonian $H_{L_1}$ defined on a finite
chain with a length $L_1$  has the following transformation property
\begin{equation}
\lim_{L_1 \rightarrow \infty} X([O_{\A}, H_{L_1}])
= \lim_{L_1 \rightarrow \infty}[X(O_{\A}), H_{L_1}]
\end{equation}
for any $X \in \G$  and for any local operator $O_{\A}$.
Thus, our model has U(1), ${\mathbb
Z}_{2}$ and translational symmetries defined by the mappings (\ref{U}),
(\ref{eq:pi}) and (\ref{T}) in the infinite-volume limit.

The spin one-point functions in the infinite-volume limit are shown
\begin{equation}
 \left.
 \begin{array}{l}
  \displaystyle
  \omega_\zeta\left(S_{x}^{(1)}\right)
  = \frac{\omega_\zeta(n_x)}{2}
  \frac{q^{x-z} + (q^{x-z})^{\ast}}{1 + |q^{x-z}|^{2}},
  \vspace*{0.5em}\\
  \displaystyle
  \omega_\zeta\left(S_{x}^{(2)}\right)
  = \frac{\omega_\zeta(n_x)}{2 i}
  \frac{q^{x-z} - (q^{x-z})^{\ast}}{1 + |q^{x-z}|^{2}},
  \vspace*{0.5em}\\
  \displaystyle
  \omega_\zeta\left(S_{x}^{(3)}\right)
  = \frac{\omega_\zeta(n_x)}{2}
  \frac{1 - | q^{x-z}|^{2}}{1 + |q^{x-z}|^{2}},
 \end{array}
 \right\}.
\end{equation}
All the symmetries of the model are spontaneously broken in the
ground state $\omega_\zeta$, since
\begin{equation}
\omega_{\zeta}( U_{\varphi}(S_{x}^{(1)}))
 \neq \omega_{\zeta}(S_{x}^{(1)}),
\quad \omega_{\zeta}(\Pi(S_{x}^{(3)})) \neq \omega_{\zeta}(S_{x}^{(3)}),
\quad \omega_{\zeta}(T_{u}(S_{x}^{(3)})) \neq \omega_{\zeta}(S_{x}^{(3)}).
\end{equation}
On the other hand, the spin one-point functions in the
all-spin-up and all-spin-down ground states are
\begin{align}
 &\omega_{ \uparrow}(S_x^{(1)}) = 0 = \omega_{ \uparrow}(S_x^{(2)}),
 \quad \quad \quad
 \omega_{ \uparrow}(S_x^{(3)}) = \frac{1}{2} \omega_{\uparrow}(n_{x}),
 \\
 &\omega_{ \downarrow}(S_x^{(1)}) = 0 = \omega_{ \downarrow}(S_x^{(2)}),
 \quad \quad \quad
 \omega_{ \downarrow}(S_x^{(3)}) = -\frac{1}{2} \omega_{\downarrow}(n_{x}).
\end{align}
Therefore, the ground states $\omega_{ \uparrow}$ and
$\omega_{ \downarrow}$ have
the U(1) and the translational symmetries.
The broken symmetry is only ${\mathbb Z}_2$ in $\omega_{ \uparrow}$ and
$\omega_{ \downarrow}$.

\subsection{The Cluster Property}

First, we give a precise definition of the cluster property.
Let $\A$ and $\B$ be finite subsets of $\Lambda$. And we
denote a translation of $\A$ by $\A + x$ for
$x \in \Lambda$:
\begin{equation}
 \A + x := \{ x + y ~|~ y \in \A \}.
\end{equation}
We say that a state $\omega$ has the cluster property, if the truncated
correlation function of arbitrary observables $O_{\A}$ and $O_{\B}$ on
arbitrary finite subsets $\A, \B \subset \Lambda$ satisfies
\begin{equation}
 \lim_{x \rightarrow \infty}
 \left[
  \omega(O_{\A} O_{\B \pm x}) -
  \omega(O_{\A}) \omega(O_{\B \pm x})
 \right]
 = 0
\end{equation}
and
\begin{equation}
 \lim_{x \rightarrow \infty}
 \left[
  \omega(O_{\A \pm x} O_{\B}) -
  \omega(O_{\A \pm x}) \omega(O_{\B})
 \right]
 = 0.
\end{equation}

Here we prove the cluster property of the ground state
$\omega_{\zeta}$. Let $O_{\A, \B}$ is a basis of local operators:
\begin{equation}
 O_{\A, \B}(\sigma_{\A}, \sigma_{\B}) =
 \left( \prod_{x \in \A} d_{x, \sigma_{x}}^{\dagger} \right)
 \left( \prod_{y \in \B} d_{y, \sigma_{y}} \right)
 \label{eq:basis_of_op}
\end{equation}
where $\sigma_{\A} = (\sigma_{x})_{x \in \A}$ with
$\sigma_{x} = \uparrow, \downarrow$.
The relation given in (\ref{eq:inv_of_c}) means that any local operator can be
written in the form
\begin{equation}
 \sum_{j} \sum_{\sigma_{\A_j}, \sigma_{\B_j}}
 \phi_{j}(\sigma_{\A_{j}}, \sigma_{\B_{j}})
 O_{\A_{j}, \B_{j}}(\sigma_{\A_{j}}, \sigma_{\B_{j}})
 \label{eq:local_op}
\end{equation}
where $\phi_{j}(\sigma_{\A_j}, \sigma_{\B_j}) \in {\mathbb C}$ is a
coefficient.
Hereafter we do not write the spin dependence explicitly for simplicity.
Note that the summation in this representation (\ref{eq:local_op})
consists of finite terms.
The cluster property is shown if we find
\begin{equation}
 \lim_{x \rightarrow \infty}
 \Bigl[
  \omega_{\zeta}(O_{\A, \B} O_{\C + x, \D + x})
  - \omega_{\zeta}(O_{\A, \B}) \omega_{\zeta}(O_{\C + x, \D + x})
 \Bigr]
 = 0, \label{eq:CP1}
\end{equation}
and
\begin{equation}
 \lim_{x \rightarrow \infty}
 \Bigl[
  \omega_{\zeta}(O_{\A - x, \B - x} O_{\C, \D})
  - \omega_{\zeta}(O_{\A - x, \B - x}) \omega_{\zeta}(O_{\C, \D})
 \Bigr]
 = 0, \label{eq:CP2}
\end{equation}
for any finite subsets $\A$, $\B$, $\C$ and $\D$ of $\Lambda$.
The following two lemmas are useful to prove these equalities: (\ref{eq:CP1})
and (\ref{eq:CP2}).

\paragraph{Lemma 1}

Let $x_{j} \in \A$ $(j = 1, 2, \cdots, |\A|)$ and
$y_{k} \in \B$ $(k = 1, 2, \cdots, |\B|)$ be elements labeled
with the monotonically increasing property
$x_{j} < x_{j + 1}$ and $y_{k} < y_{k + 1}$.
If $\langle O_{\A, \B} (\sigma_{\A}, \sigma_{\B}) \rangle_{\zeta}$ does
not vanish, then $\A$ and $\B$ satisfy the following conditions:
$\A, \B \subset \Lambda_{o}$, $|\A| = |\B|$,
$x_{j} < y_{j} \Rightarrow y_{j} < x_{j + 1}$ and
$y_{j} < x_{j} \Rightarrow x_{j} < y_{j + 1}$ for any
$ j= 1, \cdots, |\A| - 1$.
In this case, $\langle O_{\A, \B} \rangle_{\zeta}$ can be written in the
form of
\begin{equation}
 \langle O_{\A, \B} (\sigma_{\A}, \sigma_{\B}) \rangle_{\zeta}
 = \frac{B(-l, u - 1, z) B(v + 1, l, z)}{B(v + 1, l, z)}
 T_{\A, \B} (\sigma_{\A}, \sigma_{\B}),
\end{equation}
where $u = \min \{x_1, y_1\}$ and $v = \max \{ x_{|\A|}, y_{|\B|}\}$.
$T_{\A, \B}$ and its translation $T_{\A \pm w, \B \pm w}$ are finite
even in the limit of $|w| \rightarrow \infty$.

\paragraph{Lemma 2}

We consider four finite subsets $\A, \B, \C$ and $\D$ of $\Lambda_{o}$,
and define lattice sites $x = \min \A \cup \B$,
$y = \max \A \cup \B$, $u = \min \C \cup \D$ and $v = \max \C \cup \D$
with $y <  u$.
If $\langle O_{\A, \B} O_{\C, \D} \rangle_{\zeta}$ does not vanish,
we have to consider only the following two cases.
If either $\langle O_{\A, \B} \rangle_{\zeta}$ or
$\langle O_{\C, \D} \rangle_{\zeta}$ does not vanish, then
it can be written in the form:
\begin{equation}
 \langle O_{\A, \B} O_{\C, \D} \rangle_{\zeta}
 =
 \frac{B(-l, x - 1, z) B(y + 1, u - 1, z) B(v + 1, l, z)}{B(-l, l, z)}
 T_{\A, \B} T_{\C, \D}.
\end{equation}
The finite coefficient $T_{\A, \B}$ 
is identical to $T_{\A, \B} (\sigma_{\A}, \sigma_{\B})$
in Lemma 1. We have abbreviated the spin dependence.  If both
$\langle O_{\A, \B} \rangle_{\zeta}$ and
$\langle O_{\C, \D} \rangle_{\zeta}$ vanish,
the following function decays exponentially
\begin{equation}
 \langle O_{\A, \B} O_{\C, \D} \rangle_{\zeta}
 =
 \frac{B(-l, x - 1, z) B(v + 1, l, z)}{B(-l, l, z)}
 \left( -\frac{e^{i \frac{\theta}{2}}}{r} \right)^{u - y}
 \tilde{T}_{\A, \B, \C, \D}.
 \label{eq:TCF_for_vanished}
\end{equation}
$\tilde{T}_{\A, \B, \C, \D}$ and its translation
$\tilde{T}_{\A - w^{\prime}, \B - w^{\prime}, \C + w, \D + w}$ is finite
even in the limit $|w| \rightarrow \infty$,
$|w^{\prime}| \rightarrow \infty$ and both.

\paragraph{ }

We can easily prove these lemmas in practical calculations of expectation
values in the ground state $\Psi(\zeta)$ as for the density two-point
function or the electron two-point function. They will appear in
appendices \ref{A:<ee>} and \ref{A:<nn>}.

\paragraph{Proof of the Cluster Property}

Now we prove the cluster property of the ground state $\omega_\zeta$ and
excited states above $\omega_\zeta$.
In the case of $\langle O_{\A, \B} \rangle_{\zeta} = 0$, lemma 2 implies
$\langle O_{\A, \B} O_{\C, \D} \rangle_{\zeta} \rightarrow 0$ in
the limit of $(\max \A \cup \B - \min \C \cup \D) \rightarrow \infty$.
Next, we consider the case of $\langle O_{\A, \B} \rangle_{\zeta} \neq 0$
and $\langle O_{\C, \D} \rangle_{\zeta} \neq 0$.
The lemmas 1 and 2 ensure that the truncated correlation function can be
written in the form of
\begin{align}
 \langle O_{\A, \B} O_{\C, \D} \rangle_{\zeta}
 - & \langle O_{\A, \B} \rangle_{\zeta}
 \langle O_{\C, \D} \rangle_{\zeta}
 \nonumber \\
 = & T_{\A, \B} T_{\C, \D}
 \frac{B(-l, x - 1, z) B(v + 1, l, z)}{B(-l, l, z)^{2}}
 \nonumber \\
 & \times
 \left[
 B(y + 1, u - 1, z) B(-l, l, z) -
 B(-l, u - 1, z) B(y + 1, l, z)
 \right].
\end{align}
We already know that $T_{\C + w, \D + w}$ and
$\lim_{l \rightarrow \infty}B(v + w + 1, l, z)/B(-l, l, z)$ are finite
even for $w \rightarrow \infty$ and
\begin{equation*}
 \lim_{u \rightarrow \infty} \lim_{l \rightarrow \infty}
  \Bigl(
  B(y + 1, u - 1, z) B(-l, l, z) -
  B(-l, u - 1, z) B(y + 1, l, z)
  \Bigr) = 0.
\end{equation*}
This relation enables us to show the requirements (\ref{eq:CP1}) and
(\ref{eq:CP2}). Thus, the cluster property of the ground state $ \omega_{\zeta} $ has
been proved.
We can prove the cluster property of an
arbitrary excited state above the ground state $ \omega_{\zeta} $
basically in the same approach, since any excited
state can be obtained by acting some local
operators written in a finite summation of the basis
(\ref{eq:basis_of_op}) on the ground state $ \omega_{\zeta} $.
In appendix \ref{A:tran-corr-func}, we will find the following bound for a sufficiently large $u$
\begin{equation}
 \left| \lim_{l \rightarrow \infty}
  \Bigl(
  B(y + 1, u - 1, z) B(-l, l, z) -
  B(-l, u - 1, z) B(y + 1, l, z)
  \Bigr)
  \right|
  < C p^{- 2(u - w)},
  \label{eq:UB_of_TCF}
\end{equation}
with a positive number $C$
independent of $u$ for $w = \max \{ y, z \}$, $y = \max \A \cup \B$ and
$u = \min \C \cup \D$. Here, we define $p = \min \{ |q|, r \}$ for
$|q| > 1$, or $p = r$ for $|q| = 1$. This bound and the representation
(\ref{eq:TCF_for_vanished}) mean that all truncated correlation
functions decay exponentially. And their decay lengths are less than
$(2 \log p)^{-1}$.

The proof of the cluster property or other evaluations of expectation
values for the ground states $\omega_{\uparrow}$ and
$\omega_{\downarrow}$ are much easier than those for the domain wall
ground state $\omega_{\zeta}$ because the recursion relations of the
normalization functions for $\Phi_{\uparrow}$ and $\Phi_{\downarrow}$
are exactly solvable.

\setcounter{section}{0}
\setcounter{equation}{0}
\renewcommand{\thesection}{Appendix}
\renewcommand{\theequation}{\Alph{section}.\arabic{equation}}

\section{ }

Here, we show the estimates of the rescaled normalization function
to evaluate the density
one-point function and several correlation functions.
In the case of $|q| = 1$, explicit and compact representations of
expectation values are easily obtained by the exact formula of
normalization function (\ref{eq:A_with_|q|=1}).
Also in the case of $|q| \neq 1$, the dominant part of the two-point
functions is the same as those for $|q| = 1$. Here, we show the
evaluation only for $|q| > 1$ without loss of generality.

\renewcommand{\thesection}{\Alph{section}}

\subsection{Estimates of the Rescaled Normalization Function\label{A:RNF}}

Here, we show useful inequalities
\begin{align}
1 - G_{+}(x, y, z) \leq \frac{B(x, y, z)}{B(x, y + 1, z)} \leq 1,
 \quad
 1 - G_{-}(x, y, z) \leq \frac{B(x, y, z)}{B(x - 1, y, z)} \leq 1,
 \label{eq:inequ_for_B/B}
\end{align}
for $x < y$.   The positivity $1-G_{\pm}(x, y, z) > 0$
is guaranteed by the definition
\begin{align}
 G_{\pm}(x, y, z) = \min
 \left\{
 \frac{1}{r^{2} + 1}, \tilde{G}_{\pm}(x, y, z)
 \right\},
\end{align}
where the functions $\tilde {G}_{\pm}(x, y, z)$ are defined by
\begin{align}
 & \tilde{G}_{+}(x, y, z) =
 \frac{r^{-2(y - x + 1)}}{1 + r^{2}}
 \nonumber \\
 &
 + (|q|^{\frac{1}{2}} - |q|^{-\frac{1}{2}})^{2}
 \times
 \begin{cases}
 \displaystyle
  |q|^{-2 (x - z - \frac{1}{2})}
  \frac{r^{-2 (y - x + 1)} - |q|^{-2 (y - x + 1)}}{|q|^{2} - r^{2}}
  & \mbox{if $x \geq z - \frac{1}{2}$} \vspace*{1em}\\
 \displaystyle
  |q|^{- 2(z - y - \frac{3}{2})}
  \frac{1 - |q|^{-2 (y - x + 1)} r^{-2 (y - x + 1)}}{|q|^{2} r^{2} - 1}
  & \mbox{if $y \leq z - \frac{1}{2}$} \vspace*{1em}\\
 \displaystyle
  - r^{-2 (y - x + 1)}
  \frac{|q|^{-2(z - x - \frac{1}{2})}}{r^{2} |q|^{2} - 1}
  + \frac{|q|^{-2 (y - z + \frac{1}{2})}}{r^{2} - |q|^{2}}
  & \vspace*{1em}\\
  \displaystyle \quad
  + |q| r^{-2 (y - ^\prime+ 1)}
  \left(
   \frac{|q|^{2(z^\prime - z)}}{r^{2}|q|^{2} - 1}
   +\frac{|q|^{2 (z - z^\prime)}}{|q|^{2} - r^{2}}
  \right)
  & \mbox{otherwise}
 \end{cases}
 ,
\end{align}
and
\begin{align}
 & \tilde{G}_{-}(x, y, z) =
 \frac{r^{-2(y - x + 1)}}{1 + r^{2}}
 \nonumber \\
 &
 + (|q|^{\frac{1}{2}} - |q|^{-\frac{1}{2}})^{2}
 \times
 \begin{cases}
 \displaystyle
  |q|^{- 2(x - z - \frac{3}{2})}
  \frac{1 - |q|^{-2 (y - x + 1)} r^{-2 (y - x + 1)}}{|q|^{2} r^{2} - 1}
  & \mbox{if $x \geq z + \frac{1}{2}$} \vspace*{1em}\\
 \displaystyle
  |q|^{-2 (z - y - \frac{1}{2})}
  \frac{r^{-2 (y - x + 1)} - |q|^{-2 (y - x + 1)}}{|q|^{2} - r^{2}}
  & \mbox{if $y \leq z + \frac{1}{2}$} \vspace*{1em}\\
 \displaystyle
  - r^{-2 (y - x + 1)}
  \frac{|q|^{-2(y - z - \frac{1}{2})}}{r^{2} |q|^{2} - 1}
  + \frac{|q|^{-2 (z - x + \frac{1}{2})}}{r^{2} - |q|^{2}}
  & \vspace*{1em}\\
  \displaystyle \quad
  + |q| r^{-2 (z^\prime - x + 1)}
  \left(
   \frac{|q|^{2(z - z^\prime)}}{r^{2}|q|^{2} - 1}
   +\frac{|q|^{2 (z^\prime - z)}}{|q|^{2} - r^{2}}
  \right)
  & \mbox{otherwise}
 \end{cases}
 ,
\end{align}
where $z^\prime$ is an integer defined by $z^\prime \in{\mathbb Z}\cap[z-\frac{1}{2},z+\frac{1}{2})$.
These inequalities give the estimates of the expectation values.
Note that in a limit $\lambda \rightarrow \infty$, $r$ goes to
$\infty$, and therefore $B(x, y - 1, z)/B(x, y, z) \rightarrow 1$
for arbitrary $x$, $y$ and $z$.

Now we prove the inequalities (\ref{eq:inequ_for_B/B}).
The upper bounds are obvious, because the function $B(x, y, z)$ is
monotonically increasing for $y - x$.
The remaining task is to obtain the lower bounds.

Here, we treat only $B(x, y, z)/B(x, y + 1, z)$. The lower bound for
$B(x, y, z)/B(x - 1, y, z)$ can be obtained in almost the same approach.
First, we show $B(x, y, z)/B(x, y + 1, z) \geq 1 - (r^{2} + 1)^{-1}$.
Using the relation (\ref{eq:rec_of_B}), we obtain
\begin{equation}
 \frac{B(x, y, z)}{B(x, y + 1, z)}
 =
 \left[
 \frac{1 + r^{2}}{r^{2}}
 - \frac{1 - f(y + \frac{1}{2} - z)}{r^{2}}
 \frac{B(x, y - 1, z)}{B(x, y, z)}\right]^{-1} .
\end{equation}
Note $B(x, y - 1, z)/B(x, y, z) > 0$ and
$1 - f(y + \frac{1}{2} - z) > 0$. Then we obtain
\begin{equation}
 \frac{B(x, y, z)}{B(x, y + 1, z)} >
 \frac{r^{2}}{1 + r^{2}} = 1 - \frac{1}{r^{2} + 1}.
\end{equation}
To obtain another lower bound, we rewrite the recursion relation
\begin{equation}
 B(x, y + 1, z) - B(x, y, z)
 =
 \frac{1}{r^{2}}
 \left[
  B(x, y, z) - B(x, y - 1, z)
 \right]
 +\frac{f(y + \frac{1}{2} - z)}{r^{2}} B(x, y - 1, z).
\end{equation}
And using this relation recursively, we obtain
\begin{align}
 B(x, y + 1, z) - B(x, y, z)
 = &
 \frac{1}{r^{4}}
 \left[
 B(x, y - 1, z) - B(x, y - 2, z)
 \right]
 \nonumber \\
 & + \frac{f(y - \frac{1}{2} - z)}{r^{4}}
 B(x, y - 2, z)
 + \frac{f(y + \frac{1}{2} - z)}{r^{2}} B(x, y - 1, z)
 \nonumber \\
 & \vdots
 \nonumber \\
 = &
 \frac{1}{r^{2(y - x + 2)}}
 + \sum_{j = 0}^{y - x}
 \frac{f(y + \frac{1}{2} - j - z)}{r^{2(j + 1)}}
 B(x, y - j - 1, z),
\end{align}
by using $B(x, x - 2, z) = 0$.
Then we obtain
\begin{align}
 \frac{B(x, y, z)}{B(x, y + 1, z)}
 & =
 1 - \frac{1}{r^{2(y - x + 2)}} \frac{1}{B(x, y + 1, z)}
 -
 \sum_{j = 0}^{y - x}
 \frac{f(y + \frac{1}{2} - j - z)}{r^{2(j + 1)}}
 \frac{B(x, y - j - 1, z)}{B(x, y + 1, z)}
 \nonumber \\
 & \geq
 1 - \frac{r^{-2(y - x + 1)}}{1 + r^2}
 - \sum_{j = 0}^{y - x}
 \frac{f(y + \frac{1}{2} - j - z)}{r^{2(j + 1)}}
 \label{eq:deffer_of_B}
\end{align}
by using
\begin{equation}
 B(x, y + 1, z) \geq \frac{1 + r^2}{r^2}
 \quad \mbox{ and } \quad
 \frac{B(x, y - j - 1, z)}{B(x, y + 1, z)} \leq 1
\end{equation}
for $y \geq x$. Note the inequality
\begin{align}
 f(u) \leq
 (|q|^{\frac{1}{2}} - |q|^{-\frac{1}{2}})^{2}
 |q|^{- 2 |u|}.
 \label{eq:ineq_for_f}
\end{align}
We can evaluate the summation in the final line of
(\ref{eq:deffer_of_B}).
There are three cases. First case is $x \geq z - \frac{1}{2}$. Since
$y + \frac{1}{2} - j - z > 0$ for $0 \leq j \leq y - x$,
we obtain
\begin{align}
 \sum_{j = 0}^{y - x}
 \frac{f(y + \frac{1}{2} - j - z)}{r^{2(j + 1)}}
 & \leq
 \frac{(|q|^{\frac{1}{2}} - |q|^{-\frac{1}{2}})^{2}}{r^{2}}
 \sum_{j = 0}^{y - x}
 \frac{|q|^{-2 (y - j - z + \frac{1}{2})}}{r^{2 j}}
 \nonumber \\
 & =
 (|q|^{\frac{1}{2}} - |q|^{-\frac{1}{2}})^{2}
 |q|^{-2 (x - z - \frac{1}{2})}
 \frac{r^{-2 (y - x + 1)} - |q|^{-2 (y - x + 1)}}{|q|^{2} - r^{2}}.
\end{align}
For the second case $y \leq z - \frac{1}{2}$, we obtain
\begin{align}
 \sum_{j = 0}^{y - x}
 \frac{f(y + \frac{1}{2} - j - z)}{r^{2(j + 1)}}
 & \leq
 \frac{(|q|^{\frac{1}{2}} - |q|^{-\frac{1}{2}})^{2}}{r^{2}}
 \sum_{j = 0}^{y - x}
 \frac{|q|^{2 (y - j - z + \frac{1}{2})}}{r^{2 j}}
 \nonumber \\
 & =
 (|q|^{\frac{1}{2}} - |q|^{-\frac{1}{2}})^{2}
 |q|^{- 2(z - y - \frac{3}{2})}
 \frac{1 - |q|^{-2 (y - x + 1)} r^{-2 (y - x + 1)}}{|q|^{2} r^{2} - 1},
\end{align}
because $y + \frac{1}{2} - j - z < 0$ for $0 \leq j \leq y - x$.
For the third case $x < z - \frac{1}{2} < y$, we introduce an integer $n$
such that $0 \leq y - n - z + \frac{1}{2}$ and
$y - (n + 1) - z + \frac{1}{2} < 0$.
Then the summation can be evaluated by
\begin{align}
 \sum_{j = 0}^{y - x}
 \frac{f(y + \frac{1}{2} - j - z)}{r^{2(j + 1)}}
 \leq
 \frac{(|q|^{\frac{1}{2}} - |q|^{-\frac{1}{2}})^{2}}{r^{2}}
 \Biggl[ &
 \sum_{j = 0}^{n} \frac{|q|^{-2 (y - j - z + \frac{1}{2})}}{r^{2 j}}
 +
 \sum_{j = n + 1}^{y - x}
 \frac{|q|^{2 (y - j - z + \frac{1}{2})}}{r^{2 j}}
 \Biggr]
 \nonumber \\
 =
 (|q|^{\frac{1}{2}} - |q|^{-\frac{1}{2}})^{2}
 \Biggl[ &
 - r^{-2 (y - x + 1)}
 \frac{|q|^{-2(z - x - \frac{1}{2})}}{r^{2} |q|^{2} - 1}
 + \frac{|q|^{-2 (y - z + \frac{1}{2})}}{r^{2} - |q|^{2}}
 \nonumber \\
 &
 + |q| r^{-2 (y - z^\prime + 1)}
 \left(
 \frac{|q|^{2(z^\prime - z)}}{r^{2}|q|^{2} - 1}
 +\frac{|q|^{2 (z - z^\prime)}}{|q|^{2} - r^{2}}
 \right)
 \Biggr]
\end{align}
where
$z^\prime:=y-n\in{\mathbb Z}\cap[z-\frac{1}{2},z+\frac{1}{2})$.
Then we obtain another lower bound
\begin{align}
 & \frac{B(x, y, z)}{B(x, y + 1, z)}
 \geq 1 - \frac{r^{-2(y - x + 1)}}{1 + r^{2}}
 \nonumber \\
 & - (|q|^{\frac{1}{2}} - |q|^{-\frac{1}{2}})^{2} \times
 \begin{cases}
 \displaystyle
  |q|^{-2 (x - z - \frac{1}{2})}
  \frac{r^{-2 (y - x + 1)} - |q|^{-2 (y - x + 1)}}{|q|^{2} - r^{2}}
  & \mbox{if $x \geq z - \frac{1}{2}$} \vspace*{1em}\\
 \displaystyle
  |q|^{- 2(z - y - \frac{3}{2})}
  \frac{1 - |q|^{-2 (y - x + 1)} r^{-2 (y - x + 1)}}{|q|^{2} r^{2} - 1}
  & \mbox{if $y \leq z - \frac{1}{2}$} \vspace*{1em}\\
 \displaystyle
  - r^{-2 (y - x + 1)}
  \frac{|q|^{-2(z - x - \frac{1}{2})}}{r^{2} |q|^{2} - 1}
  + \frac{|q|^{-2 (y - z + \frac{1}{2})}}{r^{2} - |q|^{2}}
  & \vspace*{1em}\\
  \displaystyle \quad
  + |q| r^{-2 (y - z^\prime + 1)}
  \left(
   \frac{|q|^{2(z^\prime - z)}}{r^{2}|q|^{2} - 1}
   +\frac{|q|^{2 (z - z^\prime)}}{|q|^{2} - r^{2}}
  \right)
  & \mbox{otherwise}
 \end{cases}
 .
\end{align}

\subsection{Estimates of the Truncated Correlation Functions\label{A:tran-corr-func}}

Here, we prove the inequality (\ref{eq:UB_of_TCF}). That inequality
means that all truncated correlation functions decay exponentially.
First, we show the simple form of bounds of derivation of two rescaled
normalization function. For any $x$ and $z$,
there is a positive number $C_{0}$ independent of $y$
such that for sufficient large $y$
\begin{equation}
 \frac{B(x, y, z)}{B(x, y + 1, z)} < 1 - C_{0} p^{-2 (y + 1 - w)},
\end{equation}
with $w = \max \{ z, x \}$.
Here, we define $p = \min \{ |q|, r \}$ for $|q| > 1$, or $p = r$ for
$|q| = 1$.
We can obtain this bound easily from the bounds
(\ref{eq:inequ_for_B/B}). And following bound is trivial
\begin{equation}
 \frac{B(x, y + 1, z)}{B(x, y, z)} < 1 + C_{1} p^{-2 (y + 1 - w)}.
  \label{eq:simple_bound}
\end{equation}
Here, the finite positive number $C_{1}$ is also independent of $y$.

Let us prove the inequality (\ref{eq:UB_of_TCF}).
\begin{align}
 & B(-l, l, z) B(x + 1, y - 1, z) - B(-l, y - 1, l, z) B(x + 1, l, z)
  \nonumber \\
 & = B(-l, l, z) B(x + 1, y - 1, z)
  \left[
   1 -
   \frac{B(-l, y - 1, l, z) B(x + 1, l, z)}{B(-l, l, z) B(x + 1, y - 1, z)}
  \right]
\end{align}
We concentrate to obtain bounds of
\begin{equation}
 \frac{B(-l, y - 1, l, z) B(x + 1, l, z)}{B(-l, l, z) B(x + 1, y - 1, z)},
\end{equation}
since we have already known that the pre-factor
$B(-l, l, z) B(x + 1, y - 1, z)$ is finite even in the limit of
$l \rightarrow \infty$ and $y \rightarrow \infty$.
First we estimate an upper bound. Note that $B(x, y, z)$ is a
monotonically increasing function of $y - x$. We can evaluate
\begin{equation}
 \frac{B(-l, y - 1, l, z)}{B(-l, l, z)}
  \frac{B(x + 1, l, z)}{B(x + 1, y - 1, z)}
  < \frac{B(x + 1, l, z)}{B(x + 1, y - 1, z)}.
\end{equation}
And using Eq. (\ref{eq:simple_bound}), we obtain
\begin{align}
 \frac{B(x + 1, l, z)}{B(x + 1, y - 1, z)}
  & =
  \frac{B(x + 1, l, z)}{B(x + 1, l - 1, z)}
  \frac{B(x + 1, l - 1, z)}{B(x + 1, l - 2, z)}
  \cdots
  \frac{B(x + 1, y, z)}{B(x + 1, y - 1, z)}
  \nonumber \\
  & <
  \prod_{v = y}^{l} (1 + C_{1} p^{-2 (v - w)}).
\end{align}
To estimate this, we take the logarithm. Then we can evaluate
\begin{align}
 \log \left[ \prod_{v = y}^{l} (1 + C_{1} p^{-2 (v - w)}) \right]
  = & \sum_{v = y}^{l} \log (1 + C_{1} p^{-2 (v - w)})
  \nonumber \\
  < & C_{2} \sum_{v = y}^{l} p^{-2 (v - w)}
  = C_{3} (p^{-2 (y - w)} - p^{-2 (l - w)}),
\end{align}
for some positive numbers $C_2$ and $C_3$ independent of $y$.
Here, we have used the mean-value theorem to obtain the second line.
Then we obtain an upper bound
\begin{equation}
 \frac{B(-l, y - 1, l, z) B(x + 1, l, z)}{B(-l, l, z) B(x + 1, y - 1, z)}
  <
  e^{C_{3} (p^{-2 (y - w)} - p^{-2 (l - w)})}
   < 1 + C_{4} (p^{-2 (y - w)} - p^{-2 (l - w)})
\end{equation}
for a positive number $C_4$.
We also can evaluate a lower bound:
\begin{equation}
 \frac{B(-l, y - 1, l, z) B(x + 1, l, z)}{B(-l, l, z) B(x + 1, y - 1, z)}
  >
  1 - C_{5} (p^{-2 (y - z)} - p^{-2 (l - z)}),
\end{equation}
with a positive number $C_{5}$ independent of $y$.
Then we have obtained the bound for a sufficiently large $y$
\begin{equation}
 \left|
  \lim_{l \rightarrow \infty}
  \bigl[
  B(-l, l, z) B(x + 1, y - 1, z) - B(-l, y - 1, l, z) B(x + 1, l, z)
  \bigr]
 \right|
 < C p^{-2(y - w)}.
\end{equation}

\subsection{Estimates of the Density One-Point Function\label{A:<n>}}

Now we can estimate the density one-point function. We first estimate
the factor
\begin{equation}
 \frac{B(-l, x - 1, z) B(x + 1, l, z)}{B(-l, l, z)},
\end{equation}
which appears in the representations of the density one-point function
both for integer sites (\ref{eq:n_integer}) and half-odd-integer sites
(\ref{eq:n_h-o-i}).
Using the relation (\ref{eq:dev_of_B}), we obtain
\begin{align}
 \frac{B(-l, x - 1, z) B(x + 1, l, z)}{B(-l, l, z)}
 =
 \Biggl[
 \frac{1 + r^{2}}{r^{2}}
 - \frac{1 - f(x - \frac{1}{2} - z)}{r^{2}}
 \frac{B(-l, x - 2, z)}{B(-l, x - 1, z)}
 & \nonumber \\
 - \frac{1 - f(x + \frac{1}{2} - z)}{r^{2}}
 \frac{B(x + 2, l, z)}{B(x + 1, l, z)}
 & \Biggr]^{-1}. \nonumber
\end{align}
This form and the inequality (\ref{eq:inequ_for_B/B}) enable us to
estimate the upper bound
\begin{equation}
 \frac{B(-l, x - 1, z) B(x + 1, l, z)}{B(-l, l, z)}
 < \frac{r^{2}}{r^{2} - 1} = \frac{r}{\sqrt{\epsilon^{2} - 4}},
 \nonumber
\end{equation}
by using
\begin{equation}
 \frac{B(-l, x - 2, z)}{B(-l, x - 1, z)} \leq 1, \quad
 \frac{B(x + 2, l, z)}{B(x + 1, l, z)} \leq 1
 \nonumber
\end{equation}
and
\begin{equation}
 \textstyle 1 - f(x \pm \frac{1}{2} - z) < 1. \nonumber
\end{equation}
For the lower bound, we obtain the inequality
\begin{align}
 \frac{B(-l, x - 1, z) B(x + 1, l, z)}{B(-l, l, z)}
 > &
 \Biggl\{
 \frac{1 + r^{2}}{r^{2}}
 - \frac{1 - f(x - \frac{1}{2} - z)}{r^{2}}
 [1 - G_{+}(-l, x - 2, z)]
 \nonumber \\
 & \qquad - \frac{1 - f(x + \frac{1}{2} - z)}{r^{2}}
 [1 - G_{-}(x + 2, l, z)]
 \Biggr\}^{-1}
 \nonumber \\
 = & \frac{r}{\sqrt{\epsilon^{2} - 4}}
 \frac{1}{1 + G_{1}(x, l, z)},
\end{align}
by using
\begin{equation}
 \frac{B(-l, x - 2, z)}{B(-l, x - 1, z)} \geq 1 - G_{+}(-l, x - 2, z),
 \quad
 \frac{B(x + 2, l, z)}{B(x + 1, l, z)} \geq 1 - G_{-}(x + 2, l, z),
\end{equation}
with
\begin{align}
 G_{1} & (x, l, z) = \frac{1}{r^{2} - 1}
 \Biggl\{ f(x - \tfrac{1}{2} - z)
 [1-G_{+}(-l, x - 2, z) ]
 \nonumber \\
 & + f(x + \tfrac{1}{2} - z) [1-G_{-}(x +2,l, z) ]
+ G_{+}(-l,x - 2, z)
+ G_{-}(x + 2, l, z)
 \Biggr\}.
\end{align}

The representation of the density one-point function in terms
of the rescaled normalization function is
\begin{equation}
 \langle n_{x}\rangle_{\zeta}
 =
 \frac{\lambda^{2}}{r}
 \frac{B(-l, x - 1, z) B(x + 1, l, z)}{B(-l, l, z)}.
 \label{eq:n_integer}
\end{equation}
for $x \in \Lambda_{o}$ or
\begin{align}
 \langle n_{x} \rangle_{\zeta} =
 \frac{1}{r}
 \Biggl[
 \frac{|q|^{x - z} + |q|^{-x + z}}
 {|q|^{x - z - \frac{1}{2}} + |q|^{-x + z + \frac{1}{2}}}
 \frac{B(-l, x - \frac{3}{2}, z) B(x + \frac{1}{2}, l, z)}
 {B(-l, l, z)}
 & \nonumber \\
 \times
 \left\{
 1 - \frac{1}{r}
 \frac{|q|^{x - z} + |q|^{-x + z}}
 {|q|^{x - z + \frac{1}{2}} + |q|^{-x + z - \frac{1}{2}}}
 \frac{B(x + \frac{3}{2}, l, z)}{B(x + \frac{1}{2}, l, z)}
 \right\}
 & \nonumber \\
 +
 \frac{|q|^{x - z} + |q|^{-x + z}}
 {|q|^{x - z + \frac{1}{2}} + |q|^{-x + z - \frac{1}{2}}}
 \frac{B(-l, x - \frac{1}{2}, z) B(x + \frac{3}{2}, l, z)}
 {B(-l, l, z)}
 & \nonumber \\
 \times
 \left\{
 1 - \frac{1}{r}
 \frac{|q|^{x - z} + |q|^{-x + z}}
 {|q|^{x - z - \frac{1}{2}} + |q|^{-x + z + \frac{1}{2}}}
 \frac{B(-l, x - \frac{2}{3}, z)}{B(-l, x - \frac{1}{2}, z)}
 \right\}
 & \Biggr].
 \label{eq:n_h-o-i}
\end{align}
for $x \in \Lambda^{\prime}$.

From the representation (\ref{eq:n_integer}), we find the bounds
\begin{equation}
 \frac{\lambda^{2}}{\sqrt{\epsilon^{2} - 4}}
 \frac{1}{1 + G_{1}(x, l, z)}
 <
 \langle n_{x}\rangle_{\zeta}
 <
 \frac{\lambda^{2}}{\sqrt{\epsilon^{2} - 4}}.
\end{equation}
$G_{1}(x, l, z)$  consists of terms which decay exponentially as
$l - |x|$, $l - |z|$ or $|x - z|$ become large. And for sufficiently
large $|x - z|$, $\lim_{l \rightarrow \infty} G_{1}(x, l, z)$ is
proportional to $p^{-2|x - z|}$ asymptotically:
\begin{equation}
 \lim_{l \rightarrow \infty} G_{1}(x, l, z) = O(p^{-2|x - z|})
\end{equation}
where $p = \min\{ |q|, r \}$.

Next, we estimate the density one-point function for a site in
$\Lambda^{\prime}$. For upper bound of the density one-point function,
we evaluate
\begin{align}
 & \langle n_{x} \rangle_{\zeta}
 <
 \frac{1}{\sqrt{\epsilon^{2} - 4}}
 \Biggl[
 \frac{|q|^{x - z} + |q|^{-x + z}}
 {|q|^{x - z - \frac{1}{2}} + |q|^{-x + z + \frac{1}{2}}}
 \left\{
 1 - \frac{1}{r}
 \frac{|q|^{x - z} + |q|^{-x + z}}
 {|q|^{x - z + \frac{1}{2}} + |q|^{-x + z - \frac{1}{2}}}
 [1 - G_{-}(x + \tfrac{1}{2}, l, z)]
 \right\}
 \nonumber \\
 & \qquad \qquad +
 \frac{|q|^{x - z} + |q|^{-x + z}}
 {|q|^{x - z + \frac{1}{2}} + |q|^{-x + z - \frac{1}{2}}}
 \left\{
 1 - \frac{1}{r}
 \frac{|q|^{x - z} + |q|^{-x + z}}
 {|q|^{x - z - \frac{1}{2}} + |q|^{-x + z + \frac{1}{2}}}
 [1 - G_{+}(-l, x - \tfrac{3}{2}, z)]
 \right\}
 \Biggr]
 \nonumber \\
 & =
 \frac{1}{\sqrt{\epsilon^{2} - 4}}
 \Biggl[
 \frac{|q|^{x - z} + |q|^{-x + z}}
 {|q|^{x - z - \frac{1}{2}} + |q|^{-x + z + \frac{1}{2}}}
 \frac{
 r |q|^{x - z + \frac{1}{2}} + r |q|^{-x + z - \frac{1}{2}}
 - |q|^{x - z} - |q|^{-x + z}
 }
 {r (|q|^{x - z + \frac{1}{2}} + |q|^{-x + z - \frac{1}{2}})}
 \nonumber \\
 & \qquad \qquad \quad +
 \frac{|q|^{x - z} + |q|^{-x + z}}
 {|q|^{x - z + \frac{1}{2}} + |q|^{-x + z - \frac{1}{2}}}
 \frac{
 r |q|^{x - z - \frac{1}{2}} + r |q|^{-x + z + \frac{1}{2}}
 - |q|^{x - z} - |q|^{-x + z}
 }
 {r (|q|^{x - z - \frac{1}{2}} + |q|^{-x + z + \frac{1}{2}})}
 \nonumber \\
 & \qquad + \frac{1}{r}
 \frac{|q|^{x - z} + |q|^{-x + z}}
 {|q|^{x - z + \frac{1}{2}} + |q|^{-x + z - \frac{1}{2}}}
 G_{-}(x + \tfrac{1}{2}, l, z)
 + \frac{1}{r}
 \frac{|q|^{x - z} + |q|^{-x + z}}
 {|q|^{x - z - \frac{1}{2}} + |q|^{-x + z + \frac{1}{2}}}
 G_{+}(-l, x - \tfrac{3}{2}, z)
 \Biggr]
 \nonumber \\
 & =
 \frac{1}{\sqrt{\epsilon^{2} - 4}}
 \Biggl[
 \frac{|q|^{x - z} + |q|^{-x + z}}
 {(|q|^{x - z - \frac{1}{2}} + |q|^{-x + z + \frac{1}{2}})
 (|q|^{x - z + \frac{1}{2}} + |q|^{-x + z - \frac{1}{2}})}
 \nonumber \\
 & \qquad \times
 \Biggl\{
 |q|^{x - z}(|q|^{\frac{1}{2}} + |q|^{-\frac{1}{2}} - 2 r^{-1})
 + |q|^{-x + z}(|q|^{\frac{1}{2}} + |q|^{-\frac{1}{2}} - 2 r^{-1})
 \Biggr\}
 + G_{2}^{+ \prime}(x, l, z)
 \Biggr]
 \nonumber \\
 & =
 \frac{1}{\sqrt{\epsilon^{2} - 4}}
 \left[
 \frac{(\sqrt{\epsilon^{2} - 4} - \lambda^{2})
 (|q|^{x - z} + |q|^{-x + z})^{2}}
 {(|q|^{x - z - \frac{1}{2}} + |q|^{-x + z + \frac{1}{2}})
 (|q|^{x - z + \frac{1}{2}} + |q|^{-x + z - \frac{1}{2}})}
 + G_{2}^{+ \prime}(x, l, z)
 \right]
 \nonumber \\
 & =
 \left(1 - \frac{\lambda^{2}}{\sqrt{\epsilon^{2} - 4}}\right)
 [1 + G_{2}^{+}(x, l, z)],
\end{align}
where we have used
$|q|^{\frac{1}{2}} + |q|^{-\frac{1}{2}} = \epsilon - \lambda^{2}$, and
we determine a new function $G_{2}^{+}$, which includes
the terms which are proportional to $|q|^{-|x - z|}$.
And we obtain the lower bound
\begin{align}
 \langle n_{x} \rangle_{\zeta}
 > &
 \frac{1}{\sqrt{\epsilon^{2} - 4}}
 \Biggl[
 \frac{|q|^{x - z} + |q|^{-x + z}}
 {|q|^{x - z - \frac{1}{2}} + |q|^{-x + z + \frac{1}{2}}}
 \left\{
 1 - \frac{1}{r}
 \frac{|q|^{x - z} + |q|^{-x + z}}
 {|q|^{x - z + \frac{1}{2}} + |q|^{-x + z - \frac{1}{2}}}
 \right\}
 \frac{1}{1 + G_{1}(x - \tfrac{1}{2}, l, z)}
 \nonumber \\
 & \qquad +
 \frac{|q|^{x - z} + |q|^{-x + z}}
 {|q|^{x - z + \frac{1}{2}} + |q|^{-x + z - \frac{1}{2}}}
 \left\{
 1 - \frac{1}{r}
 \frac{|q|^{x - z} + |q|^{-x + z}}
 {|q|^{x - z - \frac{1}{2}} + |q|^{-x + z + \frac{1}{2}}}
 \right\}
 \frac{1}{1 + G_{1}(x + \tfrac{1}{2}, l, z)}
 \Biggr]
 \nonumber \\
 = &
 \frac{1}{\sqrt{\epsilon^{2} - 4}}
 \frac{
 (|q|^{x - z} + |q|^{-x + z})
 (|q|^{\frac{1}{2}} + |q|^{-\frac{1}{2}} - 2 r^{-1})}
 {(|q|^{x - z - \frac{1}{2}} + |q|^{-x + z + \frac{1}{2}})
 (|q|^{x - z + \frac{1}{2}} + |q|^{-x + z - \frac{1}{2}})}
 \nonumber \\
 & \qquad
 \times
 \left\{
 |q|^{x - z}
 \frac{1}{1+ G_{1}(x - \tfrac{1}{2}, l, z)}
 + |q|^{-x + z}
 \frac{1}{1 + G_{1}(x + \tfrac{1}{2}, l, z)}
 \right\}
 \label{eq:LB_for_Lambda-prime}\\
 = &
 \left(1 - \frac{\lambda^{2}}{\sqrt{\epsilon^{2} - 4}}\right)
 [1 - G_{2}^{-}(x, l, z)].
\end{align}
Hence we obtain bounds of the density one-point function
\begin{equation}
 \left(1 - \frac{\lambda^{2}}{\sqrt{\epsilon^{2} - 4}}\right)
 [1 - G_{2}^{-}(x, l, z)]
 < \langle n_{x} \rangle_{\zeta} <
 \left(1 - \frac{\lambda^{2}}{\sqrt{\epsilon^{2} - 4}}\right)
 [1 + G_{2}^{+}(x, l, z)]
\end{equation}
These $G_{2}^{\pm}(x, l, z)$ consist of terms which decay exponentially
for $l - |x|$, $l - |z|$ or $|x - z|$. In the infinite-volume limit,
these functions are asymptotically equal to
\begin{equation}
 \lim_{l \rightarrow \infty} G_{2}^{\pm}(x, l, z) = O(p^{-2 |x - z|}),
\end{equation}
for large $|x - z|$.
Inequality (\ref{eq:LB_for_Lambda-prime}) shows $G_{2}^{-} < 1$, and
hence the density one-point function does not vanish for all
$x \in \Lambda$.

\subsection{Estimates of the Electron Two-Point Function\label{A:<ee>}}

We can represent the electron two-point functions in the
following form
\begin{align}
 \langle c_{x, \sigma}^{\dagger} c_{y, \tau} \rangle_{\zeta}
 = & K^{\prime}(x, y, l, z)
  \frac{\eta_{x, \sigma}^{\ast} \eta_{y, \tau}}
 {\sqrt{1 + |q|^{2 (x - z)}}\sqrt{1 + |q|^{2 (y - z)}}}
 \left( -\frac{e^{i\frac{\theta}{2}}}{r} \right)^{y - x},
 \label{eq:e-e_correlation}
\end{align}
where
\begin{align}
 K^{\prime}(x, y, l, z) =
 \frac{\lambda^{2}}{r}
 \sqrt{\frac{1 + |q|^{2(y - z)}}{1 + |q|^{2(x - z)}}}
 \left(
 \prod_{w = x + 1}^{y}
 \frac{1 + |q|^{2(w - z - \frac{1}{2})}}{1 + |q|^{2(w - z)}}
 \right)
 \frac{B(-l, x - 1, z) B(y + 1, l, z)}{B(-l, l, z)}
\label{Kprime}
\end{align}
for $x, y \in \Lambda_{o}$. We can easily obtain
$K^{\prime}(x, y, l, z)$ for other $x, y \in \Lambda$, but they are
complicated forms. Since the rescaled normalization function converges
in the limit $l \rightarrow \infty$, $K^{\prime}(x, y, l, z)$ exists in
the infinite-volume limit. Apparently, $K^{\prime}(x, y, l, z)$ seems to
diverge according to the factor $\sqrt{1 + |q|^{2 (y - z)}}$. But we can
evaluate
\begin{align}
 \prod_{w = v + 1}^{y}
 \frac{1 + |q|^{2 (w - z - \frac{1}{2})}}
 {1 + |q|^{2 (w - z)}}
 = & |q|^{- (y - v)}
 \prod_{w = v + 1}^{y}
 \frac{|q|^{-2 (w - z) + 1} + 1}{|q|^{-2 (w - z)} + 1}
 \nonumber \\
 \leq & |q|^{- (y - v)}
 \prod_{w = v + 1}^{y}
 \frac{|q|^{-2 (w - z) + 2} + 1}{|q|^{-2 (w - z)} + 1}
 = |q|^{- (y - v)}
 \frac{1 + |q|^{- 2 (v - z)}}{1 + |q|^{-2 (y - z)}},
\end{align}
and then the factor
\begin{align}
 \sqrt{\frac{1 + |q|^{2(y - z)}}{1 + |q|^{2(x - z)}}} &
 \left(
 \prod_{w = x + 1}^{y}
 \frac{1 + |q|^{2(w - z - \frac{1}{2})}}{1 + |q|^{2(w - z)}}
 \right)
 \nonumber \\
 < \quad &
 \sqrt{\frac{|q|^{-2 (y - v)} + |q|^{2(v - z)}}{1 + |q|^{2(x - z)}}}
 \left(
 \prod_{w = x + 1}^{v}
 \frac{1 + |q|^{2(w - z - \frac{1}{2})}}{1 + |q|^{2(w - z)}}
 \right)
 \frac{1 + |q|^{-2 (v - z)}}{1 + |q|^{- 2 (y - v)}}
\end{align}
is finite for any $x$ and $y$. Thus, we have found that
$K^{\prime}(x, y, l, z)$ is finite for $x, y \in \Lambda_{o}$ even in
the limit of $l \rightarrow \infty$.
We can show that $K^{\prime}(x, y, l, z)$ is finite also for other
$x, y \in \Lambda$.
The expression (\ref{eq:e-e_correlation}) shows that the
electron two-point function decays exponentially as $y - x$ becomes
large with a decay length $1/\log r$ because of the finiteness of $K^{\prime}(x, y, l, z)$.
Thus, 
the electron two-point function
$\omega_{\zeta}(c_{x, \sigma}^{\dagger} c_{y, \tau})$
satisfies the
requirement of the cluster property defined in subsection
5.5.

\subsection{Results of the Density Two-Point Function\label{A:<nn>}}

We can estimate the density two-point function as well as the one-point
function.  The density two-point function is written in
\begin{equation}
 \langle n_{x} n_{y} \rangle_{\zeta}
 = \frac{\lambda^{4}}{r^{2}}
 \frac{B(-l, x - 1, z) B(x + 1, y - 1, z) B(y + 1, l, z)}{B(-l, l, z)}
 \label{eq:nn}
\end{equation}
for $x, y \in \Lambda_{o}$. We can easily obtain representation
for other $x, y \in \Lambda$, but it has a complicated form.
The representations in terms of the rescaled normalization
function give us good estimates.  From the relation (\ref{eq:dev_of_B}),
we obtain
\begin{align}
 & B(-l, l, z) =
 \left( 1 + \frac{1}{r^{2}} \right)^{2}
 B(-l, x - 1, z) B(x + 1, y - 1, z) B(y + 1, l, z)
 \nonumber \\
 & -
 \left( 1 + \frac{1}{r^{2}} \right)
 \biggl[
 \frac{1 - f(x - \frac{1}{2} - z)}{r^{2}}
 B(-l, x - 2, z) B(x + 1, y - 1, z) B(y + 1, l, z)
 \nonumber \\
 & \qquad \qquad \qquad
 + \frac{1 - f(x + \frac{1}{2} - z)}{r^{2}}
 B(-l, x - 1, z) B(x + 2, y - 1, z) B(y + 1, l, z)
 \nonumber \\
 & \qquad \qquad \qquad
 + \frac{1 - f(y - \frac{1}{2} - z)}{r^{2}}
 B(-l, x - 1, z) B(x + 1, y - 2, z) B(y + 1, l, z)
 \nonumber \\
 & \qquad \qquad \qquad
 + \frac{1 - f(y + \frac{1}{2} - z)}{r^{2}}
 B(-l, x - 1, z) B(x + 1, y - 1, z) B(y + 2, l, z)
 \biggr]
 \nonumber \\
 & +
 \frac{1 - f(x - \frac{1}{2} - z)}{r^{2}}
 \frac{1 - f(y - \frac{1}{2} - z)}{r^{2}}
 B(-l, x - 2, z) B(x + 1, y - 2, z) B(y + 1, l, z)
 \nonumber \\
 & +
 \frac{1 - f(x - \frac{1}{2} - z)}{r^{2}}
 \frac{1 - f(y + \frac{1}{2} - z)}{r^{2}}
 B(-l, x - 2, z) B(x + 1, y - 1, z) B(y + 2, l, z)
 \nonumber \\
 & +
 \frac{1 - f(x + \frac{1}{2} - z)}{r^{2}}
 \frac{1 - f(y - \frac{1}{2} - z)}{r^{2}}
 B(-l, x - 1, z) B(x + 2, y - 2, z) B(y + 1, l, z)
 \nonumber \\
 & +
 \frac{1 - f(x + \frac{1}{2} - z)}{r^{2}}
 \frac{1 - f(y + \frac{1}{2} - z)}{r^{2}}
 B(-l, x - 1, z) B(x + 2, y - 1, z) B(y + 2, l, z).
\end{align}
From this relation,
we evaluate the density two-point function
\begin{equation}
 \lim_{l \rightarrow \infty}
  \langle n_{x} n_{y} \rangle_{\zeta}
  =
  \begin{cases}
   \displaystyle
   \frac{\lambda^{4}}{\epsilon^{2} - 4}
   G_{3}(x, y, z)
   & \mbox{if } x, y \in \Lambda_{o}
   \\
   \displaystyle
   \left( 1 - \frac{\lambda^{2}}{\sqrt{\epsilon^{2} - 4}} \right)^{2}
   G_{4}(x, y, z)
   & \mbox{if } x, y \in \Lambda^{\prime}
   \\
   \displaystyle
   \frac{\lambda^{2}}{\sqrt{\epsilon^{2} - 4}}
   \left( 1 - \frac{\lambda^{2}}{\sqrt{\epsilon^{2} - 4}} \right)
   G_{5}(x, y, z)
   & \mbox{otherwise}
  \end{cases}
  .
\end{equation}
Functions $G_{3}$, $G_{4}$ and $G_{5}$ are definite
functions which approach to 1 exponentially as $|y - x|$, $|y - z|$ and
$|x - z|$ become large.

\paragraph{Acknowledgements}

The authors are grateful to T. Fujita for kindly reading the manuscript.
They would like to thank T. Koma for helpful comments.

\end{document}